\documentclass[preprint]{aastex}
\usepackage{natbib}
\setcitestyle{number}
\usepackage[utf8]{inputenc}
\usepackage{csvsimple}
\usepackage{pgfplotstable}
\usepackage{amsmath}
\usepackage{amssymb}
\usepackage{hyperref}
\usepackage{graphicx}
\usepackage{multirow}
\usepackage[draft]{todonotes}
\graphicspath{{figures_bw/}}

\newcommand{\rmd}{\mathrm{d}}
\newcommand{\ceil}[1]{\ensuremath{\left\lceil #1 \right\rceil}}

\title{The Impact of Interpixel Capacitance in CMOS Detectors on PSF shapes and Implications for WFIRST}
\shorttitle{IPC and PSFs}

\author{Arun Kannawadi}
\affil{McWilliams Center for Cosmology, Department of Physics, Carnegie Mellon University, Pittsburgh, PA 15213}
\email{arunkannawadi@cmu.edu}

\author{Charles A. Shapiro}
\affil{Jet Propulsion Laboratory, California Institute of Technology, Pasadena, CA 91109}

\author{Rachel Mandelbaum}
\affil{McWilliams Center for Cosmology, Department of Physics, Carnegie Mellon University, Pittsburgh, PA 15213}

\author{Christopher M. Hirata}
\affil{Center for Cosmology and Astroparticle Physics, The Ohio State University, 191 West Woodruff Lane, Columbus, Ohio 43210, USA.}

\author{Jeffrey W. Kruk}
\affil{NASA Goddard Space Flight Center, Greenbelt, MD 20771}

\author{Jason D. Rhodes}
\affil{Jet Propulsion Laboratory, California Institute of Technology, Pasadena, CA 91109}
\affil{California Institute of Technology, Pasadena, CA 91125}

\begin{abstract}
Unlike optical CCDs, near-infrared detectors, which are based on CMOS hybrid readout technology, typically suffer from electrical
crosstalk between the pixels. The interpixel capacitance (IPC) responsible for the crosstalk
affects the point-spread function (PSF) of the telescope, increasing the size and modifying the
shape of all objects in the images while correlating the Poisson noise.  
Upcoming weak lensing surveys that use these detectors, such as WFIRST, place stringent requirements
on the PSF size and shape (and the level at which these are known), which in turn
must be translated into requirements on IPC.   To facilitate this process, we present a first study
of the effect of IPC on WFIRST PSF sizes and shapes.  
Realistic PSFs are forward-simulated from physical principles for each WFIRST bandpass. We explore
how the PSF size and shape depends on the range of IPC coupling with pixels that are connected along
an edge or corner; for the expected level of IPC in WFIRST, IPC increases the PSF
sizes by $\sim$5\%.   We present a linear fitting formula that describes the uncertainty in
the PSF size or shape due to uncertainty in the IPC, which could arise for example due to unknown
time evolution of IPC as the detectors age or due to spatial variation of IPC across the detector.  
We also study of the effect of a small anisotropy in the IPC, which further modifies the PSF shapes.  Our
results are a first, critical step in determining the hardware and characterization requirements for the detectors used in the
WFIRST survey.
\end{abstract}
\begin{document}

\maketitle


\section{Introduction}
\label{sec:intro}


%
%
A number of ongoing and future space telescopes will focus on the near-infrared (NIR) or infrared part of the
electromagnetic spectrum, which provide a view through the gas and dust in nearby star-forming
regions, and allow higher signal-to-noise imaging of high-redshift galaxies. 
Images in the NIR are currently being taken by the Wide Field Camera 3~\citep[WFC3;][]{WFC3} in the Hubble Space Telescope and the Wide-field Infrared Survey Explorer~\citep[WISE;][]{WISE}.
Upcoming space telescope missions such as NASA's James Webb Space Telescope~\citep[JWST;][]{JWST} and Wide Field InfraRed
Space Telescope\footnote{\url{http://wfirst.gsfc.nasa.gov/}}~\citep[WFIRST;][]{WFIRST_final,spergel2013afta,spergel2015wide} will also focus on infrared imaging.
  WFIRST mission will provide high quality image data for weak gravitational lensing studies.

Weak gravitational lensing~\citep[for a review, see for example][]{2001PhR...340..291B, 2003ARA&A..41..645R, 2006glsw.conf..269S, 2008ARNPS..58...99H}
is the deflection of light rays from background sources such as
galaxies by matter in the foreground, resulting in a small magnification and shape distortion. 
Weak lensing can be a powerful cosmological probe constraining cosmological parameters~\citep[e.g.,][]{2013MNRAS.432.1544M,2013MNRAS.432.2433H,2013ApJ...765...74J,2015arXiv150705598B},
test theory of gravity on large scales~\citep[e.g.,][]{2015MNRAS.449.4326P, 2013MNRAS.429.2249S, 2010Natur.464..256R} and on smaller scales relate the galaxies to their dark matter 
halos~\citep[e.g.,][]{2012ApJ...744..159L,2013ApJ...778...93T, 2014MNRAS.437.2111V,2015MNRAS.446.1356H, 2015MNRAS.447..298H, 2015MNRAS.449.1352C,2015MNRAS.454.1161Z}.
Measuring the small but coherent distortions of the shapes of the galaxies, without the knowledge of their intrinsic shapes, 
is a challenging task that requires images with high resolution and a statistical understanding of
distances to the source galaxies~\citep[see][for recent tests on multiple  methods]{GREAT3_results1}. Weak lensing measurements in surveys like WFIRST will require detailed (sub-percent level) knowledge of the point-spread
function (PSF), from a combination of {\em a priori} modeling and empirical estimates using images of
stars, in order to remove its effect on galaxy shapes. 

While there are differences between the detectors for these missions, most notably the array size and cutoff wavelength, all of them use a hybrid CMOS readout architecture with 
mercury cadmium telluride (Hg$_{1-x}$Cd$_x$Te, often abbreviated HgCdTe) as the light-sensitive material.
In particular, they use the HxRG\footnote{HxRG stands for HAWAII x $\times$ x pixels with Reference
  pixels and Guide mode, and HAWAII stands for HgCdTe Astronomical Wide Area Infrared Imager.} family of detectors manufactured by Teledyne Imaging Sensors. 
Extensive work on understanding and characterizing the impact of detector-based effects on
astronomical imaging has been done for CCDs, which are used for optical imaging in many telescopes~\citep[see for e.g.,][and references therein]{doi:10.1117/12.672596, 1748-0221-10-05-C05026, 1748-0221-10-08-C08010}.
The use of HxRG detectors for astronomy is relatively recent, and understanding the systematic
errors in astronomical measurements that result from HxRG detector effects is an active area of research. 

Among the systematic effects in these detectors that will be relevant for WFIRST are
non-linearity, reciprocity failure~\citep{2011reciprocity}, interpixel capacitance (IPC), and persistence.
Non-linearity refers to the small non-linearity in converting the charges in the pixel to the voltage read by the read-out assembly.
Reciprocity failure usually implies that the response of a detector depends not only on the total number of incident photons, but also on how they are distributed in time
(e.g., 2$\times$ illumination for $1/2$ of the time leads to a different measurement of the total flux). This is typically a non-linear effect as well.
Persistence is the phenomenon of retaining a small but non-negligible fraction of the flux in images from the previous exposures after a reset.
Interpixel capacitance, the effect that we focus on in this study, is a form of electrical cross-talk between the pixels in the detectors. 
This is a linear effect and Sec.~\ref{sec:ipc} of this paper presents a brief review of this effect.

Any realistic galaxy image simulations used for simulating PSFs and testing weak lensing shear estimation algorithms must include these effects in the images. 
To facilitate this simulation process, the authors have built a WFIRST module within GalSim\footnote{\url{https://github.com/GalSim-developers/GalSim}}~\citep{GalSim}, an open-source simulation package, 
that has the routines to include the above mentioned detector effects and reasonable values for the parameters involved.
Additionally, values for other parameters such as the pixel scale, jitter, dark current, thermal backgrounds are also provided.
The WFIRST module also takes into account of the telescope's pupil plane configuration
and forward simulates the PSF from physical principles. 
Throughout this paper, the term ``PSF'' refers to the \emph{effective} PSF that is a convolution of the pixel response function (a 2-D top-hat function) and the optical PSF (distribution of flux at the focal plane from a point source).

The pixel scale for the WFIRST-AFTA telescope must strike a balance between the need for having a large field of view while achieving the sampling needed to resolve galaxy shapes. 
It is chosen to be $0.11''$~\citep{spergel2015wide} which causes the PSFs to be undersampled (more on this in Sec.~\ref{subsec:interleaving}).
Better sampled images are expected to be achieved in post-processing by combining the images from a dither sequence. 

The action of various detector effects cannot be expressed as a convolution for different reasons. Voltage non-linearity and reciprocity failure are intrinsically non-linear
while convolution is a linear operation. Although IPC is specified by a convolution kernel, it
cannot be considered as a part of the PSF (see Sec.~\ref{sec:implications} for details).
Both the galaxy and PSF images must be corrected for the detector effects explicitly (in
each exposure) before the galaxy image is corrected for the PSF (see, e.g., \citealt{HS03})
in order to perform shape measurement.  Any imperfections in this correction, either due to the
correction algorithm or due to insufficient knowledge of the detector effects themselves, can result
in errors in PSF models that affect weak lensing shear estimation.
To enable future studies of how detector effects and their uncertainties affect shear estimation,
here we  study the effect of IPC on the PSF images.
Thus, the goal of this study is two-fold:
\begin{enumerate}
 \item to understand how the observed PSF changes as a function of the level of coupling between
   pixels, and 
 \item to relate the errors in the coupling parameters to the errors in the reconstructed PSF.
\end{enumerate}
Ultimately, these results will be useful for setting requirements on hardware and software for the
upcoming WFIRST mission. For WFIRST, in addition to knowing the overall size of PSFs, the ability to predict them is also of utmost interest.

This paper is organized as follows. We present the detector model and explain the origin of interpixel capacitance in Sec.~\ref{sec:ipc}. 
Sec.~\ref{sec:shape} summarizes the definitions of PSF sizes and shapes used in this paper. We present in Sec.~\ref{sec:methods} the details about how we simulate the WFIRST PSFs
and how to overcome the problem of undersampling with IPC taken into account. In
Sec.~\ref{sec:results}, we present our results for how the PSF size and shape is affected by IPC.
Finally, our conclusions are summarized in Sec.~\ref{sec:conclusion}.

\section{Interpixel Capacitance}
\label{sec:ipc}
In this section, we present a brief review of interpixel capacitance, including its origins and
its effect on astronomical images.
\subsection{Detector model}
\label{sec:detector}
The photodetector array is commonly modeled (see, for example, \citealt{moore2004interpixel,moore2006quantum}) as a rectangular array of capacitors indexed by $i,j$
with node capacitance 
$C[i,j]$, each receiving a charge 
$Q[i,j]$ corresponding to the accumulated photocurrent entering the node over some integration
time. In the ideal case of zero cross talk between the nodes, the detected voltage is 
\begin{equation}
 V[i,j] = \frac{Q[i,j]}{C[i,j]}
\end{equation}

The node capacitances of all pixels are fabricated to be the same to a very high accuracy, i.e., $C[i,j] = C_\text{node}$,
making the system invariant under shifts.
If $C_\text{node}$ is a constant independent of the signal, then the photodetector array is a linear system, in addition to being shift-invariant.
This linear shift-invariant (LSI) system is characterized by a 2-dimensional impulse response function $h$. 
Thus, the voltage read out from a pixel (assuming no read noise) is given by
\begin{equation}
 V[i,j] = \left( Q * h \right)[i,j] = \sum_m \sum_n Q[m,n] h[i-m,j-n].
\end{equation}
In the ideal case, 
\begin{equation}
h_\text{ideal}[i-m,j-n] = \frac{\delta_{im}\delta_{jn}}{C_\text{node}},
\end{equation}
where $\delta_{im}$ and $\delta_{jn}$ are Kronecker delta functions. That is, the voltage in a pixel depends only on the charge accumulated in that pixel.
$1/C_\text{node}$ is then the conversion gain. Factoring out the gain explicitly gives us
\begin{equation}
 V[i,j] = \sum_m\sum_n \frac{1}{C_\text{node}} Q[m,n] K[i-m,j-n],
\end{equation}
where $K$ is called the \emph{IPC kernel}.

However, in reality, fringing fields from the edges of the node capacitors cause the voltage
readings in a pixel to depend on the charges in neighboring pixels. This effect is modeled by introducing a coupling capacitance $C_c$ between pixels.
This \emph{interpixel capacitance} in principle couples every pixel to every other pixel.  It is
different from the phenomenon of charge diffusion, which involves actual physical movement of charge
carriers to adjacent pixels; IPC is a deterministic effect arising through fringing fields without
any actual movement of charge carriers. 
CCDs are not known to exhibit any IPC, although a crosstalk due to capacitive coupling between neighboring channels in the read-out electronics exists~\citep{1748-0221-10-05-C05010}.

Pixels typically have some level of non-linear response, i.e., $C[i,j]$ varies with the charge. 
This variation of $C[i,j]$ from the nominal value of $C_\text{node}$ is modeled separately as a non-linearity
in the gain of the system, leaving IPC to be modeled as a strictly linear effect.  Identical coupling capacitances between the nodes, by fabrication, ensure that the system is still shift-variant.
\footnote{This is not strictly true. However, as long as the length scale over which the coupling changes significantly is more than a few pixels, all the arguments hold.}
Thus, the effect of interpixel capacitance can still be captured by a kernel $K$ that is convolved
with the image.

\subsection{Parametrizing IPC}
\label{sec:Parametrizing}
For any IPC kernel $K$, the entries, which refer to relative capacitance values, must satisfy $0 \le K[i,j] \le 1$ $\forall (i,j)$~\citep[see][]{moore2006quantum}.
Moreover, for unit nominal gain, the sum of all voltages is unity (in suitable units)
for a unit charge (in suitable units) in some arbitrary node. This arbitrary node, which must be far from the edges, can be defined as the origin for convenience, i.e., if $Q[m,n] = \delta_{m,0}\delta_{n,0}$, then
\begin{equation}
 C_\text{node}\sum_i\sum_j V[i,j] =1,
\end{equation}
implying
\begin{equation}
 \sum_i \sum_j K[i,j] = 1.
 \label{eq:sumK}
\end{equation}
This normalization for the IPC kernel ensures charge conservation in the case of a generic signal.

As mentioned in Sec.~\ref{sec:detector}, a capacitive coupling exists between every pair of nodes, which decreases sharply with the distance between them.
For small interpixel coupling, i.e., $C_c/C_\text{node} \ll 1$, we can consider only the coupling
between pixels sharing an edge and ignore the rest, which are typically of second or higher order in $\alpha$. In this limit, the kernel
is described by a $3\times 3$ matrix with 8 degrees of freedom (see Eq.~\ref{eq:sumK}). 
Symmetric coupling between the nodes is a reasonable assumption, i.e., $K[i,j] = K[j,i]$.
The simplest, non-trivial IPC kernel is then given by
\begin{equation}
K_\alpha = \begin{pmatrix}
           0 & \alpha & 0 \\
           \alpha & 1 - 4\alpha & \alpha \\
           0 & \alpha & 0
           \end{pmatrix}.
 \label{eq:kernel}
\end{equation}
Note that we have assumed $\alpha \ll 1$, and thus $1-4\alpha$ is always positive.

Coupling between pixels that share a corner (diagonal coupling) can be stronger than second nearest neighbor (along one of the axes) due to proximity. Thus, we can introduce 
an additional level of complexity by introducing $\alpha'$ to represent the diagonal coupling, whose value can be independent of $\alpha$.
\begin{equation}
 K_{\alpha,\alpha'} = \begin{pmatrix} \alpha' & \alpha & \alpha' \\ \alpha & 1-4(\alpha+\alpha') & \alpha \\ \alpha' & \alpha & \alpha' \end{pmatrix}
 \label{eq:twopK},
\end{equation}
where typically $\alpha' < \alpha$. 
This is a reasonable assumption since in typical H2RG devices $\alpha$ and $\alpha'$ are typically
of order $10^{-2}$ and $10^{-3}$ as we will show below.  
However, it is important to confirm that the effect of $\alpha'$ on the PSF size and shape really is
small compared to that of $\alpha$, to justify that the kernel can indeed be truncated to $3\times 3$ matrix.

There can exist a measurable asymmetry along the two axes of the detectors~\citep{hilbert2011interpixel}, i.e., 
the capacitive coupling along the rows can be different from that of the columns. Small anisotropies that arise because of this can have a significant effect on the ellipticity of objects we want to study.
This leads us to a 3-parameter kernel given by
\begin{equation}
 K_{\alpha,\alpha_+,\alpha'} = \begin{pmatrix} \alpha' & \alpha-\alpha_+ & \alpha' \\ \alpha + \alpha_+ & 1-4(\alpha+\alpha') & \alpha + \alpha_+ \\ \alpha' & \alpha - \alpha_+ & \alpha' \end{pmatrix}.
 \label{eq:threepK}
\end{equation}
Care has to be taken at the edges since the above equation cannot possibly hold. Along the edges of the physical detector, the assumed IPC model is not valid. 
But along the edges of postage stamps, one can simply extend the sky image or simply truncate the edges.



For the IPC kernel $K_\alpha$ given in Eq.~\eqref{eq:kernel}, to first order in $\alpha$, the elements of the post-IPC image $I_\text{obs}$ are
related to those of the pre-IPC image $I_\text{im}$ as 
\begin{equation}
 I_\text{obs}[i,j] = (1-4\alpha)I_\text{im}[i,j] + \alpha \left(I_\text{im}[i+1,j] + I_\text{im}[i-1,j]+ I_\text{im}[i,j+1]+I_\text{im}[i,j-1]\right).
 \label{eq:Iobs}
\end{equation}
A similar equation with more terms can be written for the IPC kernels given in Eqs.~\eqref{eq:twopK} and~\eqref{eq:threepK}.
For the IR channel of WFC3, direct measurements of the IPC kernel made
on-orbit~\citep{hilbert2011interpixel} yield 
\begin{equation}
 K_\text{WFC3} = \begin{pmatrix}
                        0.0011 \pm 0.0006 & 0.0127\pm 0.0009 & 0.0011\pm 0.0006 \\
                        0.0163 \pm 0.0014 & 0.936\pm 0.0045 & 0.0164\pm 0.0011 \\
                        0.0011 \pm 0.0006 & 0.0127\pm 0.0010 & 0.0011\pm 0.0006
                       \end{pmatrix},
\label{eq:KWFC3}                       
\end{equation}
which can be described by the 3-parameter model in Eq.~\eqref{eq:threepK}. 
For H4RGs, the nominal values for the IPC parameters (given by a subscript 0) are $\alpha_0 = 0.02$, $\alpha'_0=0.002$ and $\alpha_{+,0} = 0$ (Content, D. personal communication, 2015-11-05).

\subsection{Implications of IPC}
\label{sec:implications}
Failure to account for IPC results in 
underestimation of conversion gain~\citep{doi:10.1117/12.924968, moore2006quantum,2008SPIE.7021E..23F} and
overestimation {of various kinds of quantum efficiencies}~\citep[DQE;][]{mccullough2007measurement}.
Thus, IPC must be estimated and accounted for in order to understand the fundamental parameters of the detectors.
In addition, one must account for the effect of IPC on image shapes, which is the focus of this paper.

If one were to obtain an image of the PSF by pointing the telescope at a star, the image of the star
will include the effects of IPC. 
%
The effect of the IPC is to blur the image through a convolution. 
However, the IPC kernel may be considered as being distinct from the PSF for several reasons. Some of them are:
\begin{enumerate}
 \item The PSF is an intrinsically continuous profile that is convolved with the image and is discretized only when the light hits the detector.
       The IPC kernel, on the other hand, is inherently discrete, with a pre-defined pixel scale.
 \item The effects of IPC are centered on each pixel, independent of where the photons land.
 \item The effect of IPC comes in later than that of the PSF, at the detector level, after the
   addition of dark current, Poisson noise and nonlinearity of the conversion gain. Thus the IPC introduces correlations in the signal and noise,
whereas the PSF does not correlate the noise.
\end{enumerate}

IPC corrections are therefore different from the PSF corrections, and could be applied as a
deconvolution kernel (essentially the inverse of the IPC kernel) before the majority of the image
processing.
Moreover, they would be applied to each exposure, while PSF corrections to galaxy shape measurements
would typically be made after obtaining an oversampled image from multiple exposures.

With a precisely known IPC kernel, and in the absence of detector noise, the original image at the focal plane can be recovered by direct deconvolution.
However, read-out noise and
quantization noise is added after IPC convolution occurs.
Despite this fact, the straightforward deconvolution or division in Fourier space can be done~\citep{mccullough2008correction}, as it is stable to noise due to the absence of zeros
in the Fourier representation of the IPC kernel. This deconvolution would now introduce additional correlations in the read noise and quantization noise.
Thus an exact recovery of the noisy image, removing just the effects of the IPC, is impossible.
{As an alternative to an exact correction scheme, which would have recovered the original noisy image in the absence of noise after IPC,}
an approximate, fast correction scheme suggested in \cite{mccullough2008correction} should be evaluated to see if it is sufficient.
The approximate correction scheme suggests convolving the individual exposures with another kernel, with the sign of the IPC parameters reversed.
For the 3-parameter kernel (Eq.~\ref{eq:threepK}), it involves convolving the image with another kernel 
\begin{equation}
 K'_{\alpha,\alpha_+,\alpha'} = K_{-\alpha,-\alpha_+,-\alpha'} = \begin{pmatrix} -\alpha' & -\alpha+\alpha_+ & -\alpha' \\ -\alpha -\alpha_+ & 1+4(\alpha+\alpha') & -\alpha -\alpha_+ \\ -\alpha' & -\alpha + \alpha_+ & -\alpha' \end{pmatrix}
\end{equation}
to correct for IPC effects to first order in the coupling parameters. For the 1-parameter kernel (Eq.~\ref{eq:kernel}), 
$\left(K_\alpha \otimes K'_\alpha\right)_{ij} = \delta_{i0}\delta_{j0} + \mathcal{O}(\alpha^2)$. 
Here, the $(0,0)$ element refers to the center element of the kernel.
For $\alpha \sim 0.02$, which is roughly what H4RG detectors are anticipated to exhibit, $\alpha^2 \sim 0.0004$ which may be neglible.
For kernels with more than one parameters, bi-linear correction 
terms may exist, which are also small in magnitude.
If it turns out that $\mathcal{O}(\alpha^2)$ terms are not negligible for shape measurements for WL analysis, then one can always go for the direct deconvolution.

As stated in Sec.~\ref{sec:intro}, one of the main goals of this study is to highlight the uncertainty in the PSF  due to imperfect knowledge of the IPC parameters.
In real detectors, the coupling between the pixels varies spatially~\citep{Seshadri2008mapping}, violating the assumption that IPC is a shift-invariant effect.
However, as long as the scale on which the coupling varies is greater than a few pixels, we can treat IPC as approximately shift-invariant
with kernel parameters that vary slowly with position. Thus, the IPC parameters cannot be known
perfectly and come with errorbars, which would also be the case if there is some unknown
time-dependence of the IPC.
If $\delta\alpha$ denotes the difference between the actual parameters and their assumed (nominal) values, then $K_\alpha \otimes K'_{\alpha+\delta\alpha}$
will have terms that are of order $\delta\alpha$. 

Throughout this work, we carefully distinguish $\Delta X$ from $\delta X$; the former refers to the change in the quantity $X$ due to 
the IPC and the latter refers to the change in a quantity $X$ due to error in determining the IPC parameters or equivalently the change due to a small deviation of the IPC parameters
from their nominal values. Thus, $\Delta X$ represents a systematic change that is correctable (at least, in principle), while $\delta X$ represents a systematic error.

\section{Definitions of sizes and shapes}
\label{sec:shape}
\subsection{Based on quadrupole moments}
\label{sec:moments}
A common way to define the sizes and shapes of objects in astronomical images (PSFs, galaxies) uses weighted second moments~\citep{BJ02}.
The first moments of an image $I$ (in arbitrary units) are defined as
\begin{subequations}
\begin{equation}
{\bf x_0} = \frac{\int \rmd^2 {\bf x}\; {\bf x}\,w({\bf x})I({\bf x})}{\int\rmd^2{\bf x}\; w({\bf x})I({\bf x})},
\label{eq:first_mom}
\end{equation}
and  the second moments as
\begin{equation}
{\bf M}_{ij} = \frac{\int\rmd^2{\bf x}\;({\bf x}-{\bf x_0})_i ({\bf x}-{\bf x_0})_j w({\bf x}) I({\bf x})}
              {\int\rmd^2{\bf x}\; w({\bf x})I({\bf x})}
\label{eq:second_mom}
\end{equation}
\end{subequations}
for some weight function $w({\bf x})$. Here, ${\bf x}$ and ${\bf x_0}$ are 2-vectors, i.e., ${\bf x} = (x_1,x_2)=(x,y)$.

For a given weight function, one possible definition of linear object size $\sigma$ is given by $\left[\text{det} ({\bf M})\right]^{1/4}$.
Another is the square root of the trace of the moment matrix, $\text{tr}({\bf M})^{1/2} = \sqrt{{\bf M}_{xx}+{\bf M}_{yy}}$.
Both options have dimensions of length and are invariant under rotation; 
however, the determinant is less sensitive to the shear, so we use the determinant to define
$\sigma$.

The ellipticity of the object can be expressed in terms of the second moments as
\begin{equation}
(e_1,e_2) = \left( \frac{{\bf M}_{xx}-{\bf M}_{yy}}{{\bf M}_{xx}+{\bf M}_{yy}}, \frac{2{\bf M}_{xy}}{{\bf M}_{xx}+{\bf M}_{yy}} \right).
\end{equation}
Often, the ellipticity is expressed as a complex number ${\bf e} = e_1 + {\mathrm i} e_2$. The complex ellipticity can also be specified by the magnitude
of the ellipticity $|{\bf e}|=\sqrt{e_1^2+e_2^2}$ and an angle $\beta$, where $\beta$ is the position angle, as
\begin{equation}
(e_1,e_2) =  |{\bf e}| \left(\cos 2\beta, \sin 2\beta\right).
\end{equation}
The two linear sizes are related through the total ellipticity by the equation
\begin{equation}
 \text{tr}({\bf M}) = \frac{2\sigma^2}{\sqrt{1-|{\bf e}|^2}} .
 \label{eq:tr_sigma}
\end{equation}

\subsubsection{Choosing the weight function}
\label{sec:weight}
When $w({\bf x})$ is a constant, Eq.~\eqref{eq:second_mom} reduces to unweighted second moments,
which are divergent in the presence of noise~\citep{KSB95}. Moreover, diffraction-limited PSFs are
Airy-like, with intensity decreasing with distance from the image center as $1/r^3$ for large $r$. Thus, the elements in the moment matrix
for an Airy PSF diverge logarithmically and hence are formally infinite. In practice, due to the finite size of the detector, one
would obtain finite values, but they would depend
strongly on the number of pixels used to calculate the moments, which is undesirable. This is true independent of the exact form of the PSF when the pupil of the telescope
has a sharp edge.

\cite{KSB95} introduced circular Gaussian weight functions in order to obtain finite values of higher order moments. \cite{BJ02} generalized
the weight function to be an elliptical Gaussian that matches the shape of the object.
This can be achieved in principle by finding the best-fit elliptical Gaussian to the image by minimising
\begin{equation}
 E = \int \rmd^2 {\bf x} \; \left| I({\bf x}) - A\exp\left[ -\frac{1}{2}({\bf x}-{\bf
       x_0})^\text{T}{\bf M}^{-1}({\bf x}-{\bf x_0})\right]\right|^2 \label{eq:energy}
 \end{equation}
over the six independent variables in $(A, {\bf x}_0, {\bf M})$. The optimal values of ${\bf x}_0$ and ${\bf M}$ satisfy Eqs.~\eqref{eq:first_mom} and~\eqref{eq:second_mom}
for the weight function
\begin{equation}
 w({\bf x}) = \exp\left[ -\frac{1}{2}({\bf x}-{\bf x}_0)^\text{T}{\bf M}^{-1} ({\bf x}-{\bf x}_0) \right].
\end{equation}
In practice, it is more common to determine this weight function via an iterative process than by
minimising Eq.~\eqref{eq:energy}, resulting in the term ``adaptive moments''.

For Gaussian objects, the adaptive and unweighted moments are equivalent, while the adaptive
size of a non-Gaussian PSF is typically smaller than its unweighted size since the former downweights the
extended wings of the PSF.

\subsubsection{Transformation properties}
\label{sec:transformation}
The image arising from the convolution of the IPC kernel with a given image can be seen as the sum of shifted and scaled versions of the original image (see Eq.~\ref{eq:Iobs}).
Eq.~\ref{eq:Iobs} is a particular case of a generic linear transformation of the image given as
\begin{equation}
 I({\bf x}) \rightarrow I_\text{new}({\bf x}) = (K_\text{IPC}\otimes I)({\bf x}) = \sum_{\rmd{\bf x}} \lambda(\rmd{\bf x}) I({\bf x}+\rmd{\bf x}).
 \label{eq:transformation}
\end{equation}
for some set of coefficients $\lambda$.
The transformation of the best-fit Gaussian parameters (obtained in the form of adaptive moments) under the individual operations of rescaling
and shifting are simple: 
\begin{equation}
   I({\bf x}) \rightarrow I_\text{new}({\bf x}) = \lambda I({\bf x}) \implies \left( A, {\bf x}_0, {\bf M }\right) \rightarrow \left( \lambda A, {\bf x}_0, {\bf M }\right)
  \end{equation}
  \begin{equation}
     I({\bf x}) \rightarrow I_\text{new}({\bf x}) = I({\bf x}+\Delta{\bf x}_0) \implies \left( A, {\bf x}_0, {\bf M }\right) \rightarrow \left( A, {\bf x}_0 + \Delta {\bf x}_0, {\bf M }\right)
 \end{equation}
A combined spatial translation and intensity rescaling of an image is trivial. However, the operation of finding the best-fit single Gaussian does not behave in any simple way under a general linear transformation,
i.e., the best-fit Gaussian to an image $I_1({\bf x})$ and the best-fit Gaussian to an image $I_2({\bf x})$ do not determine the best-fit Gaussian to an image $I_3({\bf x}) = I_1({\bf x}) + I_2({\bf x})$ {in a straightforward manner}.


Unlike the adaptive moments, the unweighted moments are amenable to analytical calculations.
Since the unweighted moments of two images simply add to give the unweighted moments of the third image obtained by convolving the two images,
we can write an expression for the unweighted size (if it exists) of \emph{any} object as a function of the two IPC coupling parameters $\alpha$ and $\alpha'$,
including the pixel response as
\begin{align}
 \sigma_\text{un,obs}(\alpha,\alpha') = \sqrt{\sigma_\text{un}^2 + 2\alpha + 4\alpha'} = \sqrt{\sigma_\text{un,int}^2 + \frac{1}{12}+2\alpha+4\alpha'},
 \label{eq:exact_unweighted_size}
\end{align}
where $\sigma_\text{un,obs}$ and $\sigma_\text{un}$ are the unweighted sizes with and
  without the effects of IPC, and
$\sigma_\text{un,int}$ is the intrinsic size of the object (without the broadening due to the pixel response) in pixels. We derive this relation in Appendix~\ref{app:unweighted_mom}.

The expression derived may not be  useful for space-based PSFs, but for larger objects $(\sigma_\text{un,int} \gg \sqrt{1/12})$ that are approximately Gaussian (for which the adaptive and unweighted moments agree),
it can serve as a good approximation for the adaptive size if $\alpha' < \alpha \ll 1$. The
conditions on $\alpha$, $\alpha'$, and $\sigma_\text{un,int}$ ensure that the 
non-Gaussianity in the final image is small. As an expression for the unweighted size, Eq.~\eqref{eq:exact_unweighted_size} holds true in all cases (if $\sigma$ exists).

\subsection{Other size definitions}
\label{sec:nonmoment}
There are other size definitions that are not based on second moments.
The full width at half maximum (FWHM) is one common measure of the size of a PSF. 
For an Airy PSF corresponding to a wavelength $\lambda$,
\begin{equation}
 \text{FWHM (in pixels)}  \approx \frac{1.03 \lambda}{sD},
 \label{eq:FWHM}
\end{equation}
where $D$ is the telescope diameter and $s$ is the pixel scale in units of radians/pix. 

Another 
measure of size is the half-light radius, $R_e$, which is the radius of the circle
centered at the object centroid that encloses 50\% 
of the energy. It is sometimes denoted as $EE50$.  For an Airy PSF,
\begin{equation}
 R_e = EE50 \text{  (in pixels)  } \approx 0.535\frac{\lambda}{sD}.
 \label{eq:EE50}
\end{equation}
Note that the $EE50$ is a radius and the FWHM is more like a diameter. 

Diffraction-limited PSFs have large wings. The adaptive moments represent the core size of the PSF,
while the unweighted moments and (to a lesser extent) the non-moment-based sizes
also capture information about these wings. The WFIRST Science Definition Team report~\citep[SDT;][]{spergel2015wide} specifies the requirement of PSF sizes in terms of $EE50$.
However, for weak lensing systematics, we focus on the PSF core size from the adaptive moments.
For comparison, we present the half-light radius and adaptive size for the PSFs in Table~\ref{tab:size_hlr}, with and without diffraction
spikes (which were neglected in the WFIRST SDT report).

Realistic PSFs have features such as diffraction spikes from the supporting struts, central obscuration, aberrations and
pixelization and hence the resulting profile is not an Airy pattern 
(see Fig.~\ref{fig:PSFs}). Thus, one cannot use Eq.~\ref{eq:FWHM} or Eq.~\ref{eq:EE50} to estimate
the PSF size. 
One way to define these quantities is to fit the image to a profile with well-defined FWHM or
$EE50$.  However, as we will see in Sec.~\ref{sec:methods}, the WFIRST PSFs have complex features that are not captured by simple PSF models.
Thus, we will use an empirical measurement of $EE50$ directly from the images.

Measuring FWHM directly from the PSF image is a difficult task and is highly sensitive to noise in the image.
Measuring the half-light radius is comparatively robust since we integrate (partially) the light profile instead of using the individual samples of the profile.
The first task in measuring the half-light radius of a PSF is to identify the centroid of the image.
Starting from that location, we find the radius of the smallest circle that contains at least 50\% of the total flux.
This is done by calculating the distance of every pixel from the center and determining empirically the azimuthally averaged radial profile.
That is, given a center $(x_0,y_0)$ and a separation from the centroid $R$, the fraction of light enclosed $f(R;x_0,y_0)$ is 
\begin{equation}
 f(R;x_0,y_0) = \sum_{i,j} I[i,j] \Theta(R^2-(i-x_0)^2-(j-y_0)^2),
\end{equation} 
where $\Theta()$ is the Heaviside step function.
Here we have implicitly assumed that the PSF has unit flux.
The half-light radius is calculated by solving for $R$ such that $f(R;x_0,y_0) = 0.5$ for some $x_0,y_0$.

The half-light radius can be fairly sensitive to the choice of the centroid. 
For Airy-like PSFs that exhibit circular symmetry, which would be the case if the supporting struts were radial (see Sec.~\ref{subsec:wfirst} for details)
and in the absence of aberrations, the center is unambiguously the peak of the profile.
But the WFIRST PSFs do not exhibit any such symmetry and hence it is not evident where the centroid should be.
In simulations, one can locate the `\emph{true}' center by identifying the center of the underlying Airy profile and
treat aberrations as re-distribution the light around the `true' center, thus affecting the half-light radius (but not the centroid itself).
However, in real observations of stars to determine the PSF, it is impossible to know where the `true' centroid would be and thus we need a prescription to calculate the centroid
given a PSF image.
One natural choice is to use the coordinates obtained from Eq.~\ref{eq:first_mom} as the centroid for calculating the half-light radius, which will not coincide with
the `true' center of the PSF in general.

\section{Methods}
\label{sec:methods}
\subsection{WFIRST module description}
\label{subsec:wfirst}
Realistic WFIRST PSFs used in this work are simulated using GalSim~\cite{GalSim}.
GalSim is a commonly-used open-source
tool for the weak lensing community to simulate realistic images of galaxies.
The authors of this paper have built a WFIRST module within GalSim, which is now publicly available in
GalSim v1.3. We provide a brief description of the WFIRST module below.

Filter responses for the six bandpasses - Z087, Y106, J129, W149, H158 and F184 - are available in this module, along with the blue and red wavelength limits for each of the filters
and the corresponding effective wavelengths, which are the bandpass-weighted mean wavelength. This information, listed in Table~\ref{tab:band_description}, can be retrieved by calling the \texttt{getBandpasses} routine within the WFIRST module.

The focal plane assembly (FPA) of the WFIRST-AFTA telescope consists of 18 H4RG detectors.
PSFs corresponding to one or more of these detectors can be generated by passing the detector numbers as parameters to the argument \texttt{SCAs} of \texttt{getPSF}.
The variations of the PSFs within each detector were verified to be small and GalSim v1.3 does not account for this variation, though future releases of GalSim may include them.

Complex, aberrated wavefronts incident on a circular pupil can be approximated by a sum of Zernike polynomials. 
Following the convention in \cite{Noll76}, the polynomials are labelled by an integer $j$. 
The low order Zernike polynomials map the low-order aberrations commonly found in telescopes, such as defocus $(j=4)$, astigmatism $(j=5,6)$, coma $(j=7,8)$,
trefoil $(j=9,10)$, and so on. 
Using these polynomials up to $j=11$, corresponding to spherical aberration, the WFIRST PSFs
corresponding to any of 18 detectors can be constructed from the \texttt{getPSF} routine.

By default, \texttt{getPSF} outputs instances of the \texttt{ChromaticOpticalPSF} class which can be
convolved with a galaxy or star that has an SED to make a chromatic image.
Alternatively, the user has an option to obtain instances of achromatic \texttt{OpticalPSF} class, which evaluates the chromatic WFIRST PSFs
at a particular wavelength provided by the user. For the results described in this paper, PSFs for each of the six bandpasses are obtained by evaluating the chromatic PSFs at the corresponding effective wavelengths.

\begin{figure}
 \centering
 \includegraphics[scale=0.5]{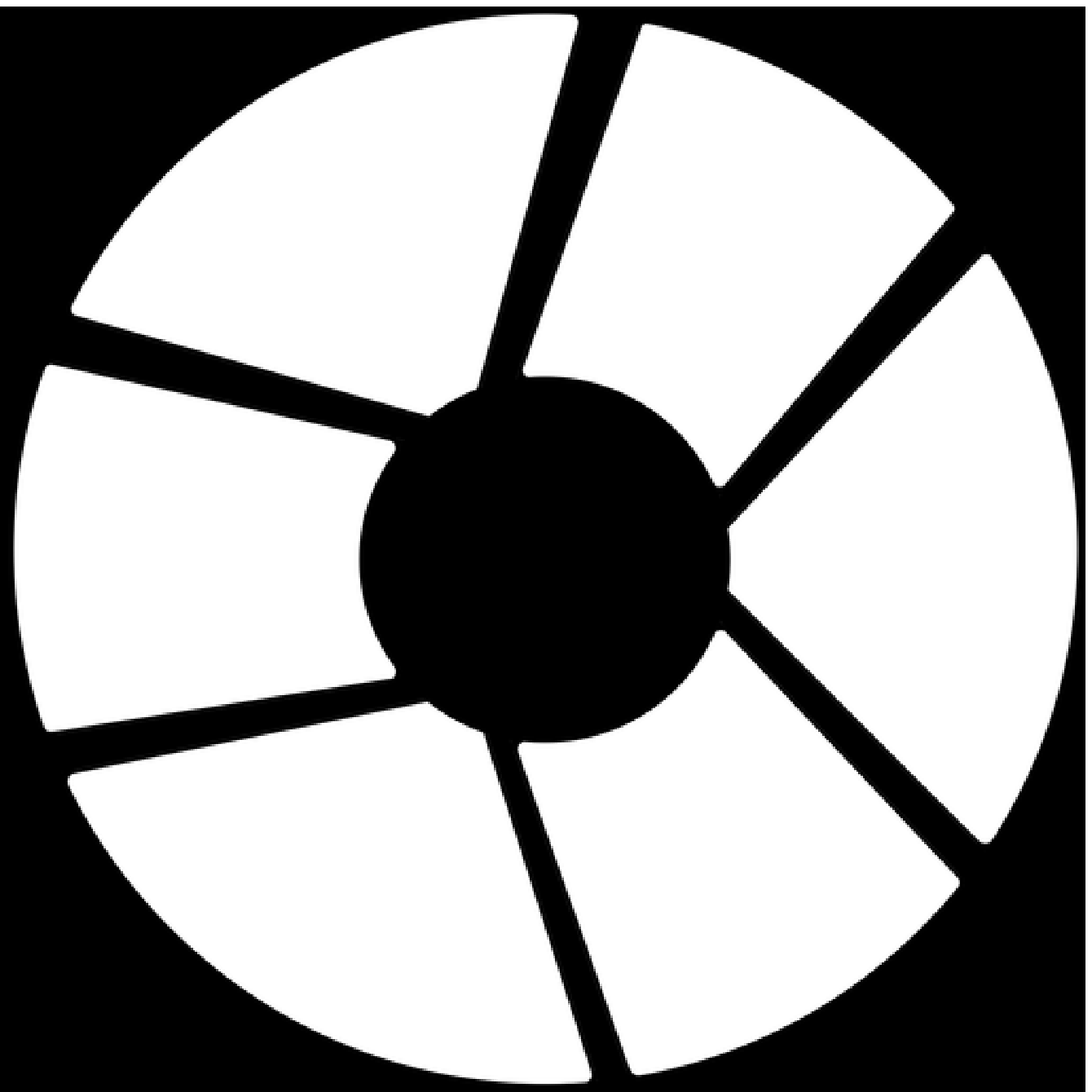} \hfill
 \includegraphics[scale=0.5]{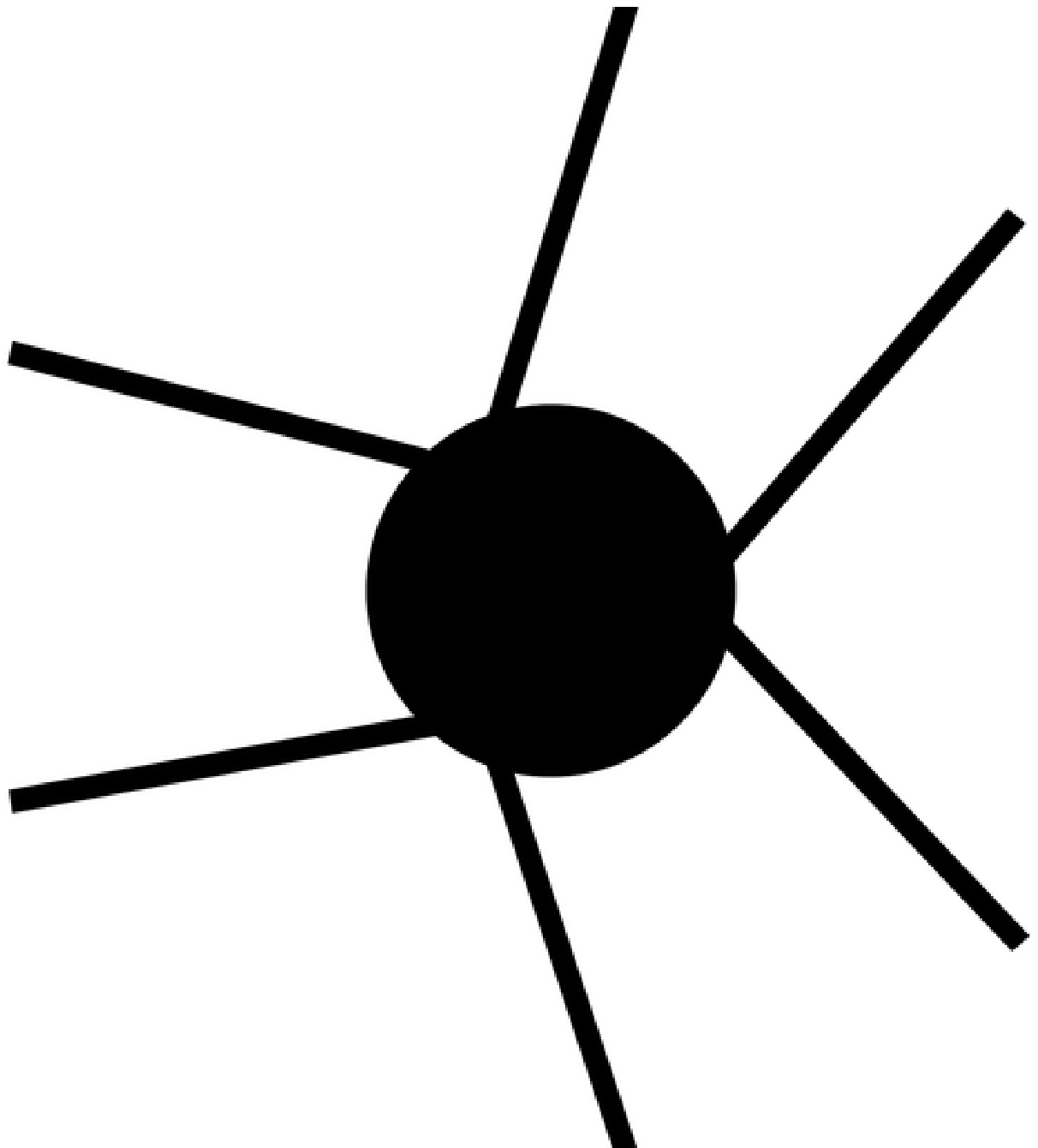}
 \caption{{\em Left:} WFIRST pupil plane configuration for long wavelength bands W149, H158, and
   F184. {\em Right:} WFIRST pupil plane configuration
 for short wavelength bands Z087, Y106, and J129. The simulations in this work use the pupil plane
 image on the left for all wavelengths.}
 \label{fig:pupil_plane}
\end{figure}

The WFIRST-AFTA telescope has a central obscuration in the pupil plane and the supporting struts are not radial (see Fig.~\ref{fig:pupil_plane}).
The \texttt{getPSF} routine in the WFIRST module takes the pupil plane configuration into account and simulates the PSF images, with expected aberrations according
to the latest design  with chromatic effects if
requested by the user, for each of the bandpasses and for each of the 18 H4RG detectors
in the focal plane~\citep{pasquale2014optical}. This process
requires two inputs: 
the pupil plane configuration and the aberrations described by the Zernike coefficients, which
are available publicly \footnote{\url{http://wfirst.gsfc.nasa.gov/science/sdt\_public/wps/references/instrument/}}. The latter have been incorporated within the WFIRST module
but in GalSim v1.3, only the circular pupil plane configuration, which is appropriate for the long-wavelength bands, is incorporated.
%

\begin{center}
\begin{figure}
\centering
 \includegraphics[width=0.32\columnwidth]{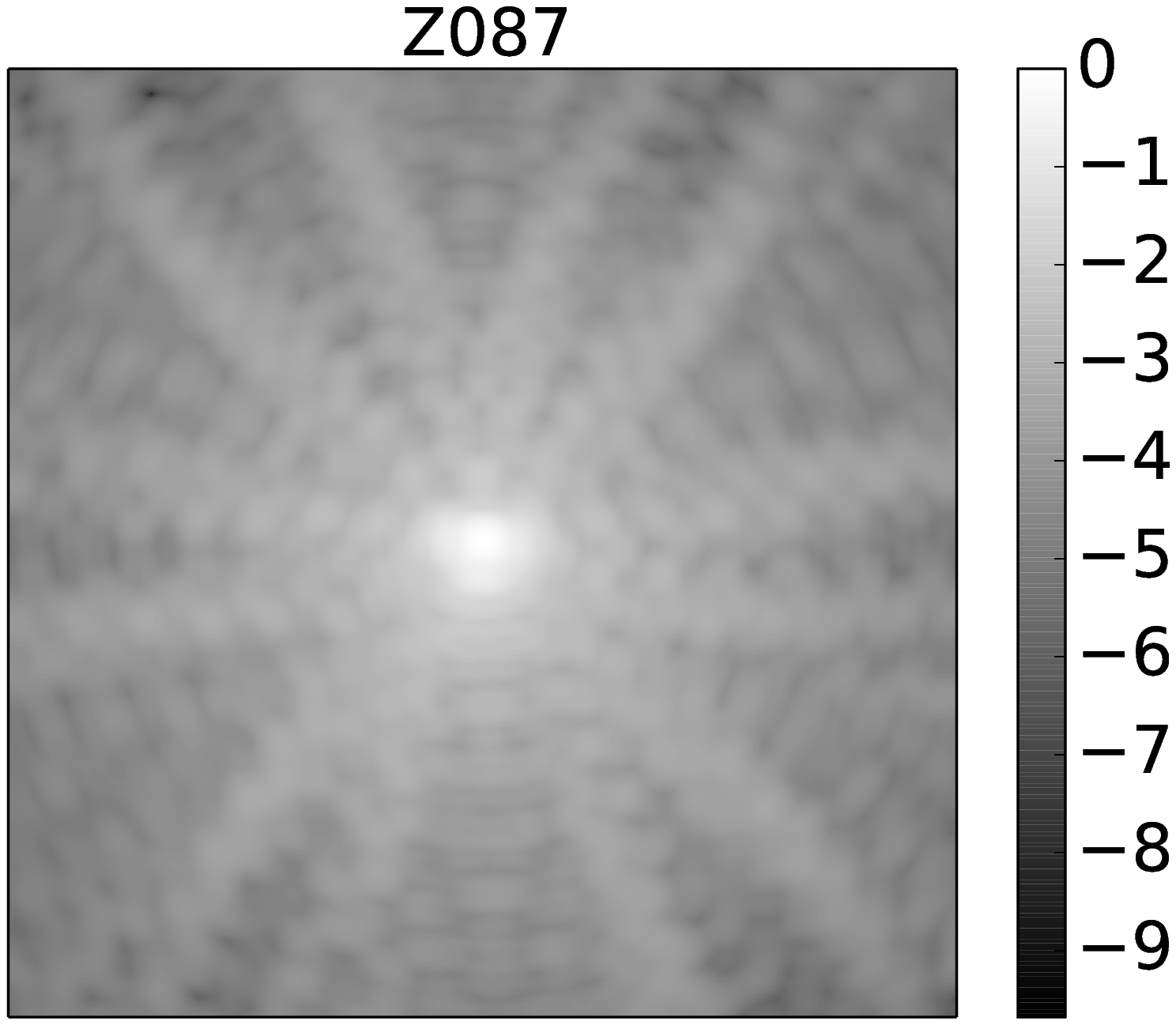}
\includegraphics[width=0.32\columnwidth]{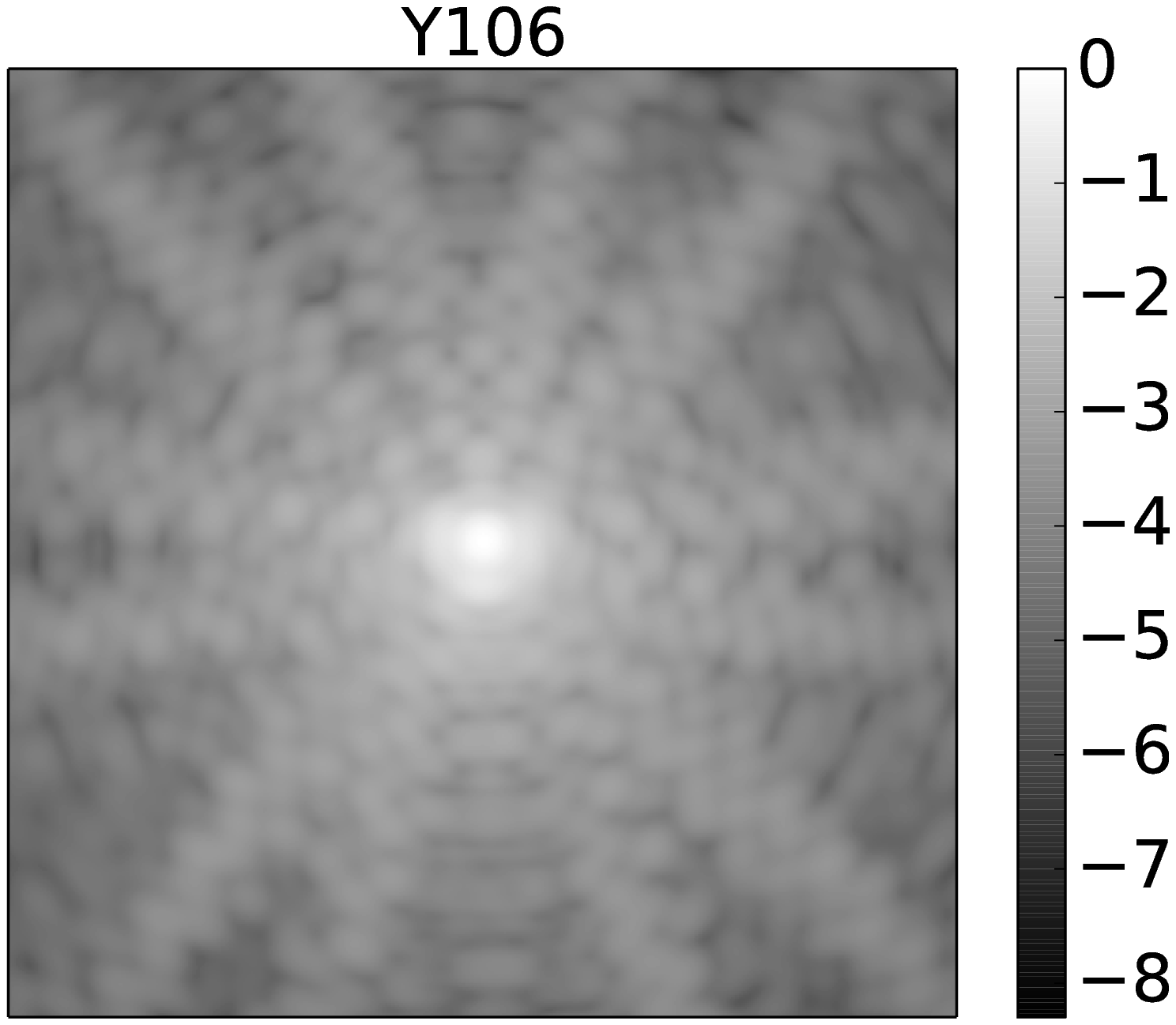}
\includegraphics[width=0.32\columnwidth]{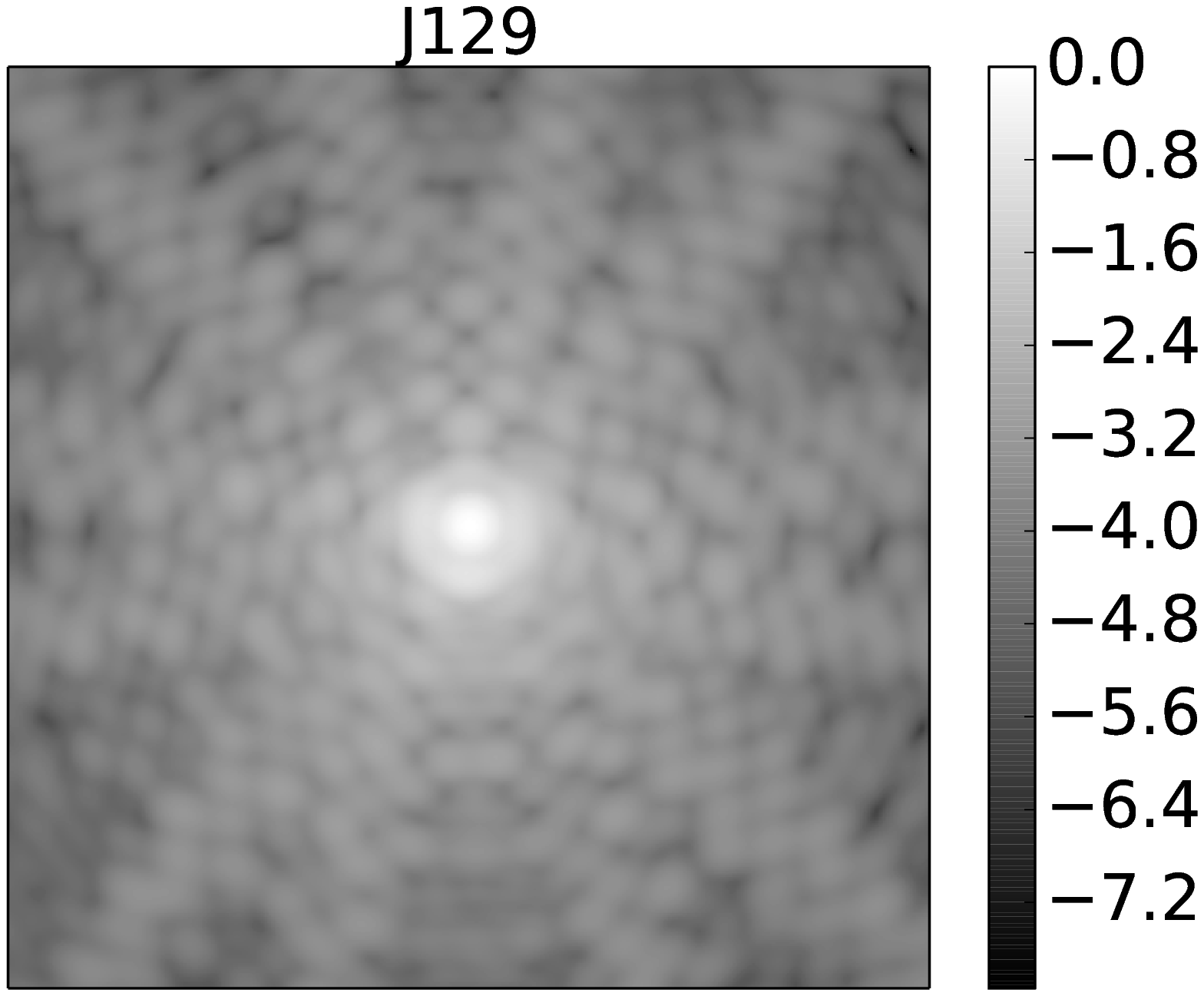}\\
\includegraphics[width=0.32\columnwidth]{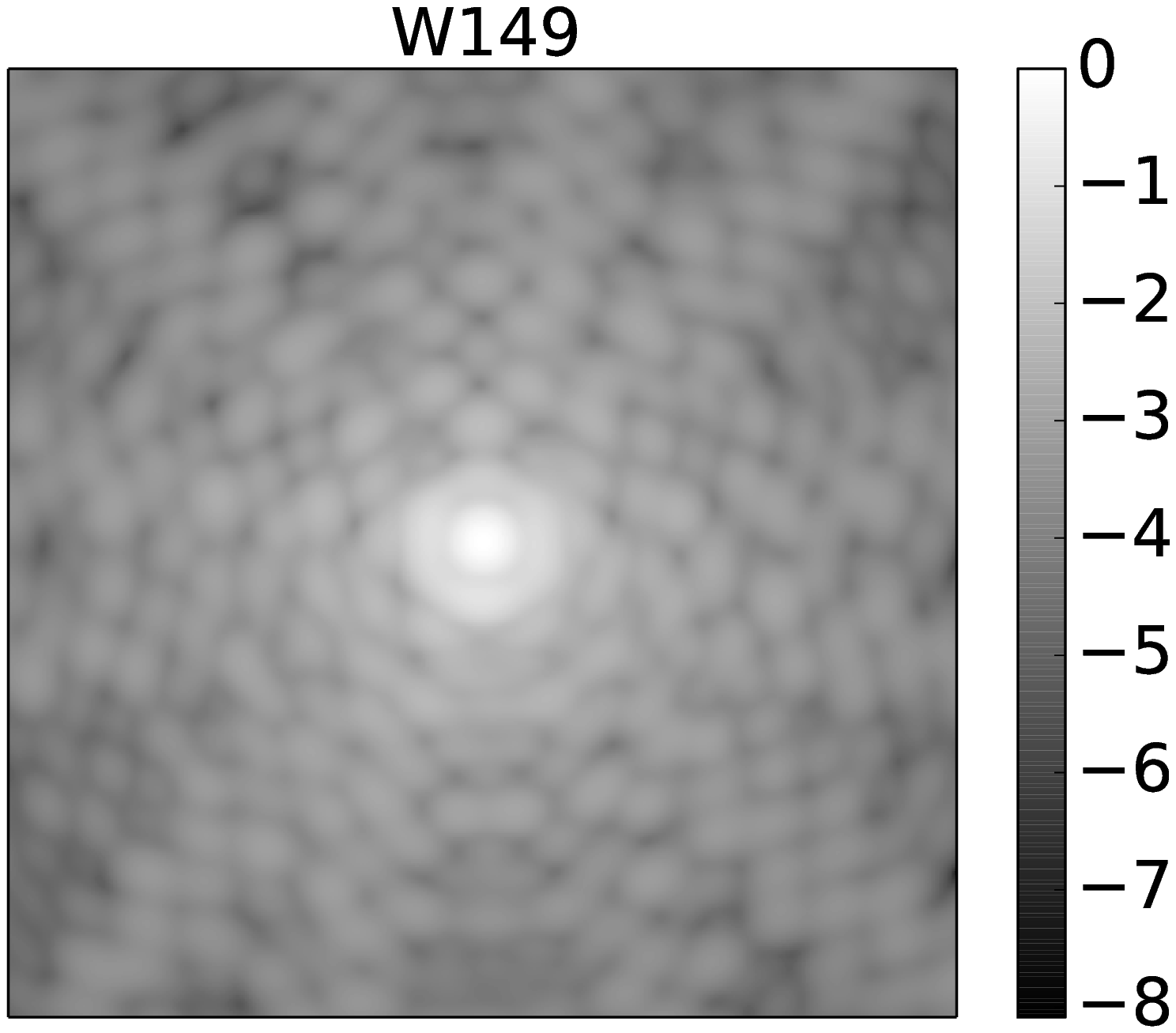}
\includegraphics[width=0.32\columnwidth]{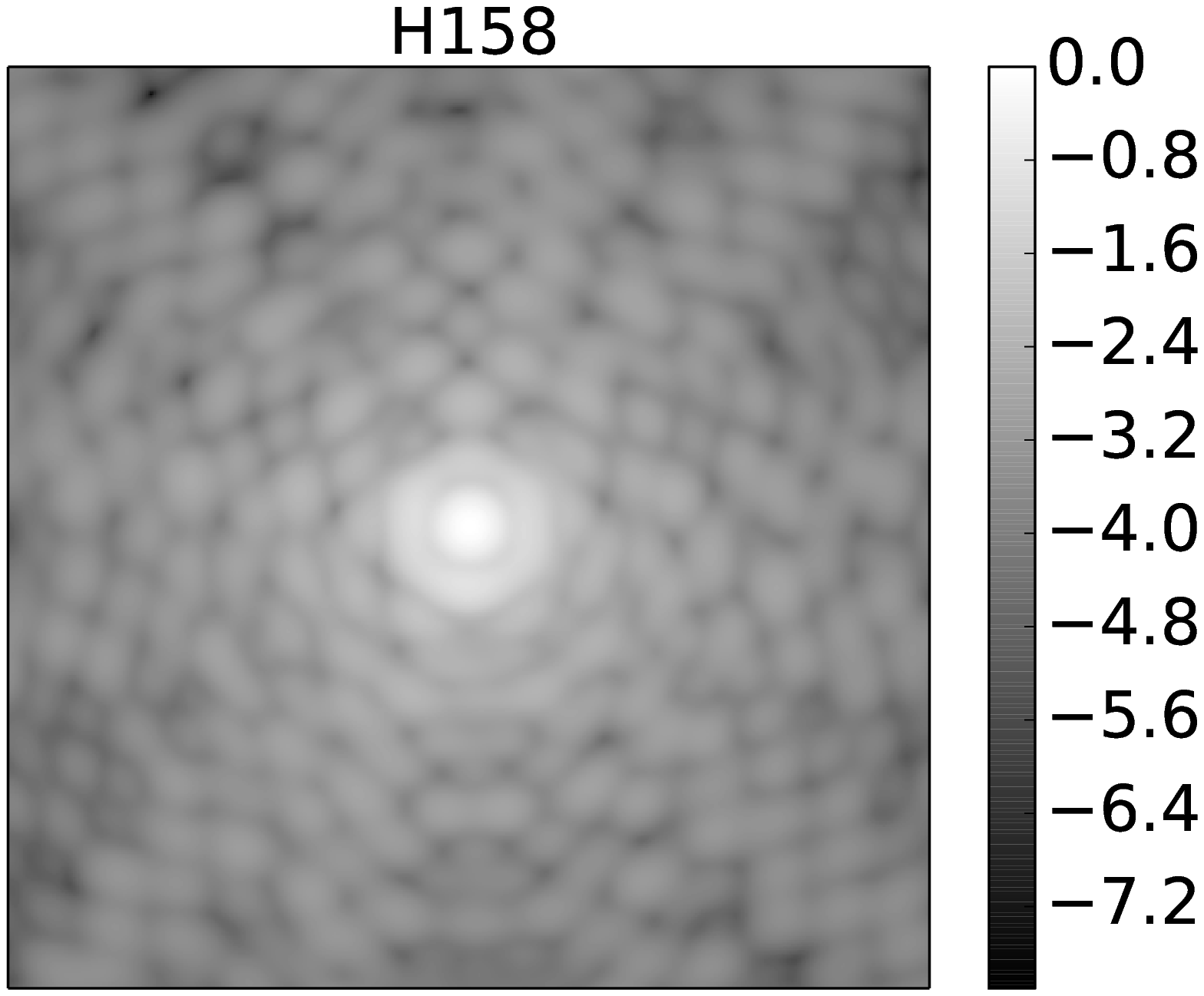}
\includegraphics[width=0.32\columnwidth]{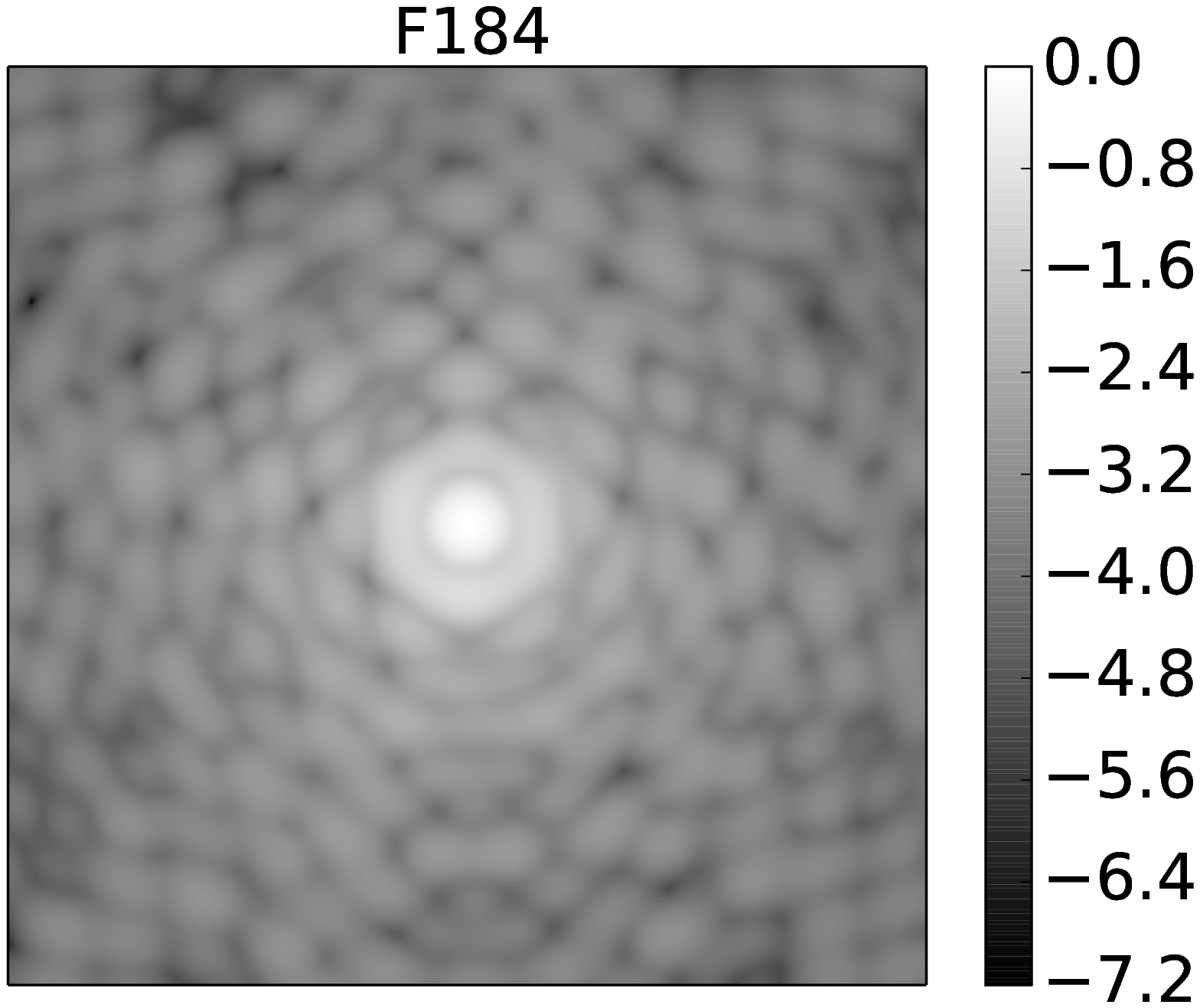}
 \caption{Simulated PSF images using the WFIRST module in GalSim for different filters, in logarithmic intensity scale, relative to the brightest pixel at the center. The PSFs are evaluated at the effective wavelength
 of the bandpass filters listed in Table.~\ref{tab:size_hlr}. The images correspond to an area of $1.76''\times1.76''$ or $16 \times 16$ pixels, drawn at a scale
 that is 32 times smaller than the native pixel scale, thus giving images of $512 \times 512$ pixels.
 The filters are arranged in increasing wavelengths from left to right, top to bottom, i.e.,
 the top row corresponds to  Z087, Y106, and J129 (from left to right) and the bottom row
 corresponds to W149, H158, and F184 bands (from left to right).}
 \label{fig:PSFs}
\end{figure}
\end{center}

The PSFs can be drawn as a GalSim \texttt{Image} instance with the (approximate) native WFIRST pixel scale of $0.11$\arcsec\ per
pixel (\texttt{wfirst.pixel\_scale}), which results in realistically undersampled images. 
The resolution of a PSF image can be increased by convolving the PSF profile with a top-hat profile corresponding to the pixel response and then drawing the image
at a smaller scale. Fig.~\ref{fig:PSFs} shows higher-resolution images of the PSFs generated in such a manner in the absence of IPC or any other detector effects and noise.
However, resampling the PSF image in the presence of IPC requires combining multiple dithered exposures containing IPC. 
We explain in Sec.~\ref{subsec:interleaving} how to handle PSF images in the presence of IPC.


The WFIRST module also has routines to incorporate many detector non-idealities such as
nonlinearity, reciprocity failure and IPC, and to model the detector configuration for an
observation at a given position with a given orientation angle. 
Values for telescope parameters such as the  pupil diameter and obscuration; detector parameters
such as the pixel scale, dark current, and IPC coupling; and basic survey parameters like planned
bandpasses and exposure times  
are also included. 
These values will be adjusted as the WFIRST design evolves.
\begin{table}
 \centering
 \begin{tabular}{c|c|c|c}
  \hline
  Bandpass & Minimum $\lambda$ & Maximum $\lambda$ & Effective $\lambda$ \\
  \hline
  Z087 & 735.0 & 1010.0 & 873.39 \\
  Y106 & 900.0 & 1230.0 & 1061.43 \\
  J129 & 1095.0 & 1500.0 & 1292.11 \\
  W149 & 905.0 & 2050.0 & 1458.01 \\
  H158 & 1340.0 & 1830.0 & 1577.05 \\
  F184 & 1630.0 & 2060.0 & 1837.3 \\
  \hline
 \end{tabular}
 \caption{Table of minimum, maximum and effective wavelengths in nanometers for each of the six bandpasses.}
 \label{tab:band_description}
\end{table}

\subsection{Simulation }
To study the effects of IPC on PSFs in the simplest possible setting, we ignore all other detector effects (non-linearity, reciprocity, read noise etc.).
Effects that are non-linear will lead to slightly different PSF images depending on the apparent magnitude of the star and the exposure time; ignoring them helps
to understand the effects on PSFs independent of such quantities.
We also add no noise in our simulations since that would require specification of a particular signal level.

Resampling the PSF profile (see Sec.~\ref{subsec:interleaving}) to obtain high resolution PSF images ensures that our results are less sensitive to 
where the centroid of the PSF falls within a pixel (center or edge or corner).  
We let the signal level of the PSF images be arbitrary since we do not include noise or any signal-dependent effects, and since IPC is linear it does not depend on the signal level.
For all the reasons mentioned above, it suffices to simulate one PSF image per band for each of the detectors.

The results presented below are for the center of a randomly chosen detector, $7$~\citep[see][for numbering system]{pasquale2014optical}. We confirmed that our results
are not significantly altered when using other detectors.
The WFIRST reference weak lensing program requires shape measurements in only the J129, H158 and F184 bands. Thus, we present our results only for the PSFs of these 3 bands.

It is important to understand what features are included in our PSF simulations and what features are not.
First, the PSFs include the diffraction spikes
due to the supporting struts, while the WFIRST SDT report set requirements while masking out these
spikes.  
The WFIRST requirements~\citep{spergel2015wide} also considered jitter and charge diffusion, which are not included in our simulations. 
Thus, while the simulated PSFs are fairly realistic, they differ in certain ways from the PSFs on
which the requirements were set. Also, the short-wavelength bands - Z087, Y106 and J129 - do not use
the appropriate pupil plane configuration shown in Fig.~\ref{fig:pupil_plane}, but rather use that of the long-wavelength bands
throughout this work.

\subsection{Overcoming undersampling}\label{subsec:interleaving} 

Measurements of object sizes or shapes from undersampled images can significantly differ depending on where the object centroid 
falls within a pixel. The  image sampling rate must be at least the Nyquist rate
for the band limit set by the optical response of the system in order 
to recover the full continuous image from discrete pixel values, and thereby avoid aliasing. 
We want to analyse only oversampled images to distinguish detector effects from aliasing artifacts.

WFIRST PSFs are not Nyquist sampled, by design. In order to measure the change in the PSF sizes due to
interpixel capacitance, we must increase the resolution of the PSF image. In GalSim, the sampling rate of the PSF image can be increased
when calling the \texttt{drawImage} routine with a scale parameter \texttt{scale=$s/N$} for some $N>1$, with $s$ being the native pixel scale.
This will alter the pixel response as well and hence does not produce the correct PSF image.
One must first convolve the PSF with the pixel response corresponding to the native pixel scale, then call \texttt{drawImage} with \texttt{method=`no\_pixel'}.
For detector effects like voltage non-linearity and reciprocity failure \citep[see,
e.g.,][]{2011reciprocity} for which the detector effect  depends on the pixel value, the above
method of obtaining Nyquist-sampled images is sufficient.

However, to include interpixel capacitance, for which the pixel correction depends on neighboring pixel values, images 
must be drawn at the native pixel scale.
Higher-resolution images can be obtained from multiple lower-resolution images with sub-pixel
offsets, also known as a dither sequence. Softwares like Drizzle~\citep{Drizzle},
iDrizzle~\citep{iDrizzle}, and 
IMCOM~\citep{IMCOM_algo} use algorithms to combine dither sequences. However, for uniform sub-pixel offsets, the resolution can be increased
by simply interleaving the images - a technique that dates back to \cite{1999Lauer}. This technique is ideal for PSF images from simulations, where offsets
can be precisely set, allowing  higher sampling to be achieved without external (to
GalSim) image combination software.

The basic interleaving concept is mathematically described as follows: 
Let $I_{n\times n}[\cdot\;,\;\cdot]$ denote some  $n \times n$ image of some continuous light
profile $I(x,y)$ hitting the detector.  Thus
\begin{equation}
 I_{n\times n}[i,j] = \int_{-\frac{p}{2}}^{\frac{p}{2}}\rmd x'\;\int_{-\frac{p}{2}}^{\frac{p}{2}}  \rmd y'\; I(ip+x',jp+y') \;\;\; (i,j\in \left\lbrace 1,2,\dots,n\right\rbrace),
\end{equation}
where $p$ is the pixel spacing or more appropriately, the length of the side of a pixel in the above equation.
Consider a set of 3 images $I^{(m)}_{n\times n}[\cdot\;,\;\cdot]$ for $m=1,2,3$, 
obtained by moving the detector by a distance $p/2$ along either of the axes or both, with
$I^{(0)}_{n\times n} = I_{n\times n}$. Therefore
\begin{equation}
 I^{(m)}_{n \times n}[i,j] = \int_{-\frac{p}{2}}^{\frac{p}{2}}\rmd x'\;\int_{-\frac{p}{2}}^{\frac{p}{2}}  \rmd y'\; I\left((i+\frac{b_1}{2})p+x',(j+\frac{b_2}{2})p+y'\right),
\end{equation}
with $b_1, b_2 \in \left\lbrace 0, 1 \right\rbrace$ and $m = 2b_1 + b_2$. Given these four images,
one can obtain a $2n\times 2n$ image of $I(x,y)$ with twice the resolution than the original  by simply interleaving these 4 images:
\begin{equation}
 I_{2n\times 2n}[2i+b_1,2j+b_2] = I^{(m)}_{n\times n}[i,j],
 \label{eq:interleave}
\end{equation}
This image has an effective sampling interval of $p/2$ while the pixel response function is still $p$. Note that the interleaved image has a higher flux given by the sum of the fluxes
of the individual images. Since we are interested in measuring only the second moments, the normalization does not matter.
\begin{equation}
 I_{2n\times 2n}[i,j] = \int_{-\frac{p}{2}}^{\frac{p}{2}}\rmd x'\;\int_{-\frac{p}{2}}^{\frac{p}{2}}  \rmd y'\; I(ip/2+x',jp/2+y') \;\;\; (i,j\in \left\lbrace 1,2,\dots,2n\right\rbrace)
 \label{eq:smaller_pixel_scale}
\end{equation}
One can interleave $N\times N \; (N\ge 2)$ images in either direction
to generate an image with a sampling rate that is $N$ times greater than the native one, without changing the pixel response.
The \texttt{interleaveImages} routine in \texttt{galsim.utilities} carries out this process for
GalSim users.

An oversampled image of the WFIRST PSF with detector effects included must be 
obtained by interleaving multiple undersampled PSF images that have these effects included.
For band-limited PSFs, one can in principle reconstruct the Nyquist-sampled PSF from a `superimage' of the PSF by
choosing $N = \ceil{(2p)/(\lambda_\text{min}N_f)}$, where $\lambda_\text{min}$ is the smallest wavelength in a given bandpass filter
and $N_f$ is the focal ratio or the f-number of the telescope. For the WFIRST-AFTA telescope, $p=10\mu m$ and $N_f = 7.8$~\citep{spergel2015wide}.
This gives $N \ge 3$ for J129 and $N \ge 2$ for H158 and F184 bands.
However, we choose to oversample the images by a large factor in order to additionally reduce the quantization error,
the error due to discretizing, and to reduce the variance due to any sub-pixel offsets that PSFs may have.
We find that setting $N=32$ is sufficient for all our analysis by checking for convergence in our results as a function of $N$.

The oversampled image is a special linear combination of the undersampled images, such as IMCOM
would produce from a dither sequence. 
Oversampling images  in this manner does not introduce any shape measurement error \citep{IMCOM_WLsystematics}.

\section{Results}
\label{sec:results}
\subsection{Increase in PSF sizes}
 \label{subsec:changes}
 \begin{table}
\centering
\begin{tabular}{c|cc|cc|cc}
 \hline
 \multirow{2}{*}{Bandpass} 
 & \multicolumn{2}{c}{Adaptive size $\sigma$ (in mas)} & \multicolumn{2}{c}{HLR $R_e$ (in mas)} & \multicolumn{2}{c}{Ellipticity $|e|$} \\
 & w/ spikes & w/o spikes & w/ spikes & w/o spikes & w/ spikes & w/o spikes \\
 \hline
 J129  & 61.50 & 60.480 & 114.85 & 95.06 & 0.0646 & 0.0652 \\
 H158 & 66.16 & 65.884 & 123.72 & 103.85 & 0.0379 & 0.0417 \\
 F184 & 72.39 & 72.452 & 132.37 & 113.21 & 0.0246 & 0.0292 \\
 \hline
 \end{tabular}
 \caption{Table of PSF sizes and ellipticity. Adaptive sizes and half-light radii in milliarcseconds (mas) and magnitude of ellipticities are tabulated for the PSFs of the J129, H158 and F184 bandpasses.
  Size and ellipticity measurements for the PSFs without including the diffraction spikes are also tabulated,
  so as to be able to relate to the values in WFIRST documents~\cite{spergel2013afta,spergel2015wide}.}
 \label{tab:size_hlr}
\end{table}

The High Latitude Imaging Survey Data Set requirements on the half-light radii of the PSFs are: $\le 0.12''$ for the J129 band, $\le 0.13''$ for the H158 band and $\le 0.14''$
for the F184 band.  These requirements were set using specific methodology, including the masking of
diffraction spikes and inclusion of jitter and charge diffusion.  We cannot compare our PSF sizes
with those from \cite{spergel2015wide} due to these methodology differences.  However, since our
goal is to determine how IPC affects PSF sizes, we provide our estimates of PSF sizes without IPC (computed
various ways) in Table~\ref{tab:size_hlr} as a baseline, then consider increases with respect to
that baseline. 

As mentioned in Sec.~\ref{sec:nonmoment}, the half-light radius depends on the choice of the centroid and 
the centroid computed from Eq.~\ref{eq:first_mom} will in general not agree with the `true' center of the image.
The centroid from the adaptive moments disagrees with the `true' center of the image by at most half a pixel ($0.055''$) in either direction.
Prior to the application of IPC, the half-light radius determined using Eq.~\ref{eq:first_mom} for
the centroid is smaller compared to the one calculated using the `true' center of the image by $2-4\%$. 

\begin{figure}
 \centering
 \includegraphics[width=\columnwidth]{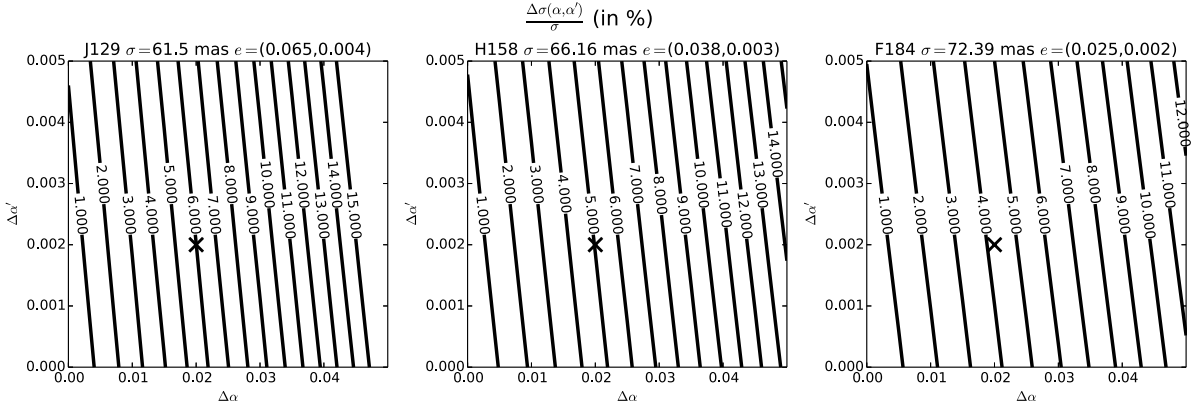}
 \caption{Contour plot of the relative increase in the adaptive size (expressed as a percent) as a
   function of the $\alpha$ and $\alpha'$ parameters in the IPC kernel in Eq.~\eqref{eq:twopK} for the relevant WFIRST PSFs. The black $\times$ represents a nominal value of the IPC parameters in H4RG detectors.
 For each filter, the adaptive size in milliarcseconds (mas) and ellipticity $(e_1, e_2)$ without IPC is noted above the
 subplots.} 
 \label{fig:accu_sigma}
\end{figure}

\begin{figure}
 \centering
 \includegraphics[width=\columnwidth]{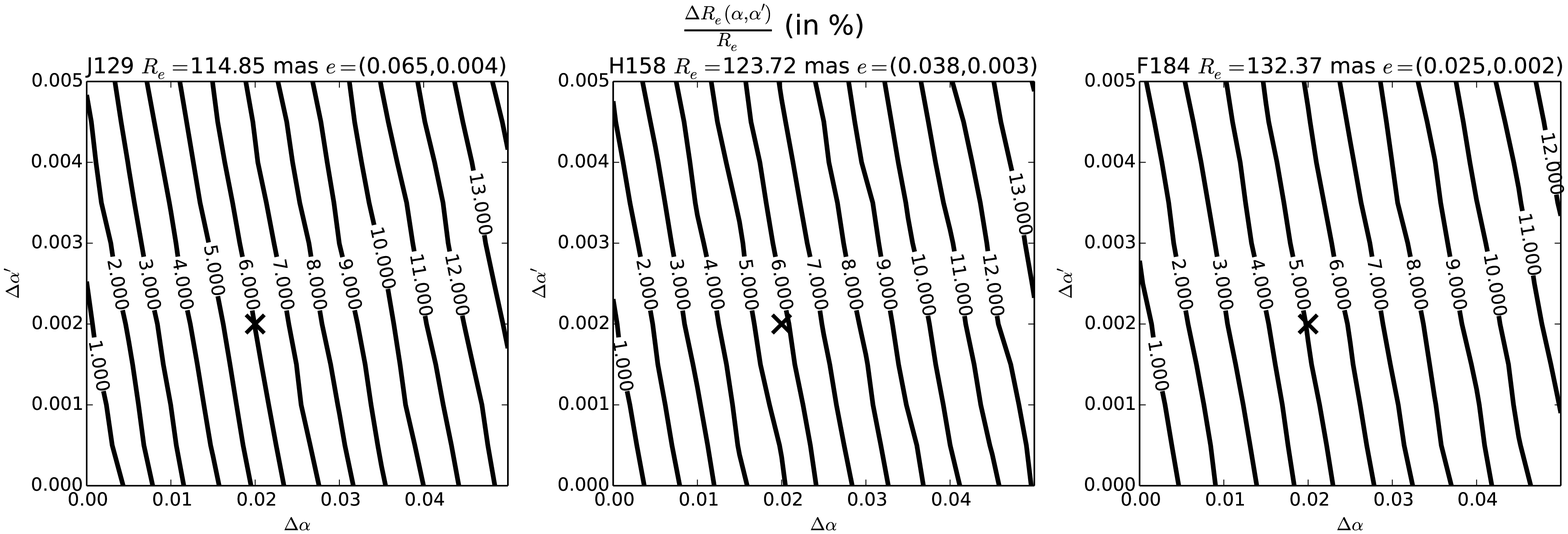}
 \caption{Contour plot of the relative increase in the half-light radius EE50 (expressed as a percent) as a
   function of the $\alpha$ and $\alpha'$ parameters in the IPC kernel in Eq.~\eqref{eq:twopK} for the relevant WFIRST PSFs. The black $\times$ represents a nominal value of the IPC parameters in H4RG detectors.
 For each filter, the EE50 in milliarcseconds (mas) and ellipticity $(e_1, e_2)$ without IPC is noted above the
 subplots. The black cross marker represents a nominal value of the IPC parameters in the H4RG detectors.}
 \label{fig:accu_hlr}
\end{figure}

We present our results for two specific sub-classes of the 3-parameter IPC kernel (Eq.~\ref{eq:threepK}). 
The first of them is the 2-parameter isotropic IPC model (Eq.~\ref{eq:twopK}). We consider this case since the nominal value of $\alpha_+$ is 0.
Figures~\ref{fig:accu_sigma} and~\ref{fig:accu_hlr}
show the increase in PSF size in all bands as a function of $\alpha$ and $\alpha'$, using the
adaptive size (Fig.~\ref{fig:accu_sigma}) and the EE50 size (Fig.~\ref{fig:accu_hlr}). 
As shown in Figs.~\ref{fig:accu_sigma}-~\ref{fig:accu_hlr}, the increase in PSF sizes is a few per cent in all bands for the nominal values ($\alpha_0=0.02$ and $\alpha'_0=0.002$) of the IPC parameters.
More specifically, the relative change in the adaptive size is about $4-6\%$ and the relative change in the half-light radius (with Eq.~\ref{eq:first_mom} as centroid)
is about $5-6\%$ depending on the bandpass. 
When the half-light radius is computed with the `true' center as the centroid, the relative increase is smaller $(\sim 4.5\%)$
and the variation is smaller across the bandpasses. 
We emphasize that it is impossible to know the location of the `true' center in the real observations of PSF and thus results with `true' center being the centroid are not relevant in practice.
We nevertheless present these results to show that not knowing the `true' center does not significantly affect our overall conclusions.

\begin{figure}
 \centering
  \includegraphics[width=\columnwidth]{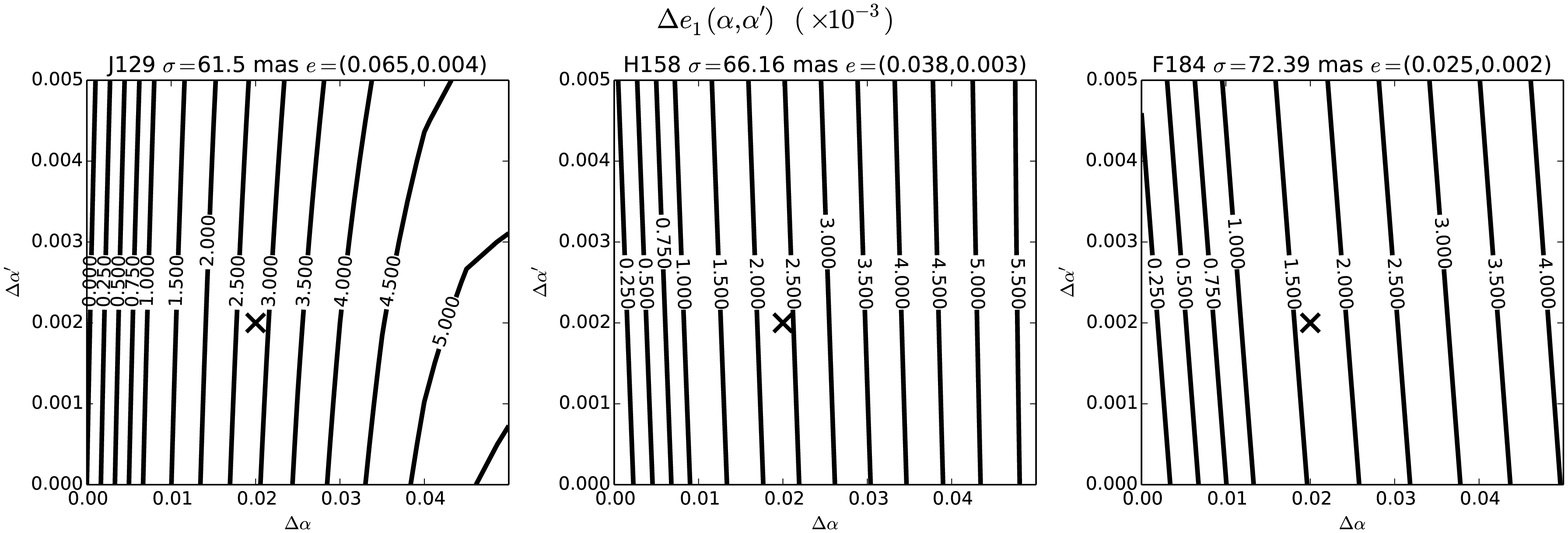}
  \caption{Contour plot of the change in the ellipticity component $e_1$ as a
   function of the $\alpha$ and $\alpha'$ parameters in the IPC kernel in Eq.~\eqref{eq:twopK} for the relevant WFIRST PSFs. The black $\times$ represents a nominal value of the IPC parameters in H4RG detectors.
 For each filter, the adaptive size in milliarcseconds (mas) and ellipticity $(e_1,e_2)$ without IPC is noted above the subplots.}
  \label{fig:accu_e1}
\end{figure}
\begin{figure}
 \centering
 \includegraphics[width=\columnwidth]{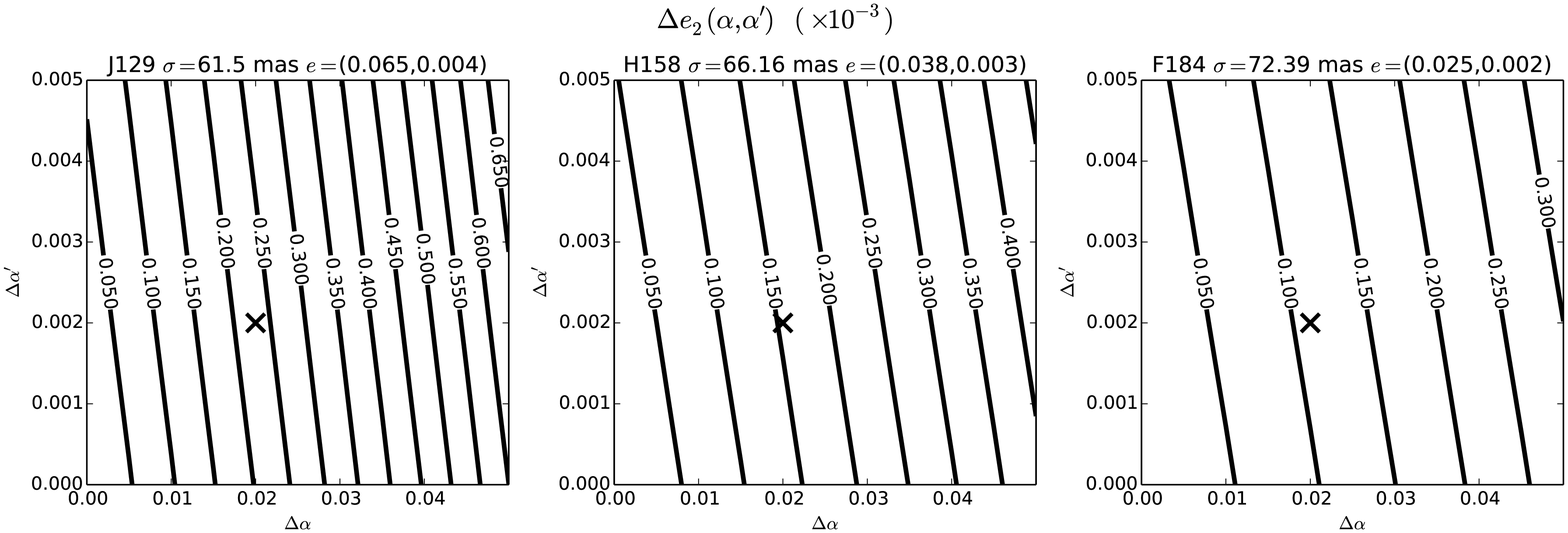}
 \caption{Contour plot of the change in the ellipticity component $e_2$ as a
   function of the $\alpha$ and $\alpha'$ parameters in the IPC kernel in Eq.~\eqref{eq:twopK} for the relevant WFIRST PSFs. The black $\times$ represents a nominal value of the IPC parameters in H4RG detectors.
 For each filter, the adaptive size in milliarcseconds (mas) and ellipticity $(e_1,e_2)$ without IPC is noted above the subplots.}
 \label{fig:accu_e2}
\end{figure}

As expected, the size is affected more by the coupling between nearest neighbors than by diagonal coupling, owing to a higher coupling parameter.
The same is true for the change in PSF ellipticity shown in 
Figs.~\ref{fig:accu_e1} and~\ref{fig:accu_e2} for the ellipticity components $e_1$ and $e_2$, along
and at $45^\circ$ with respect to the pixel edges respectively. A typical value for $\Delta e_1$
($\Delta e_2$) is
$\sim 10^{-3}$ ($10^{-4}$). Thus, even if uncorrected, 
the contribution of IPC to the PSF anisotropy and hence to the additive bias in the shear estimate is expected to be small.

\begin{figure}
 \centering
 \includegraphics[width=\columnwidth]{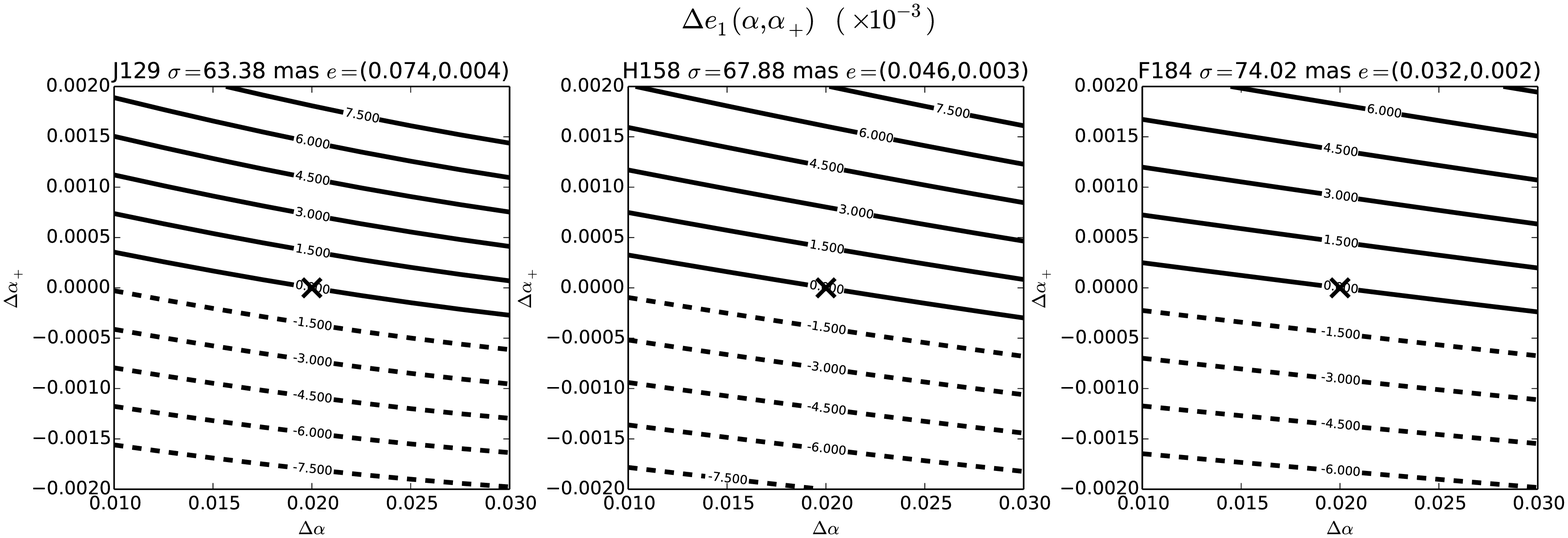}
 \caption{Contour plot of the change in the ellipticity component $e_1$ as a
   function of the $\alpha$ and $\alpha_+$ parameters in the IPC kernel in Eq.~\eqref{eq:threepK} with $\alpha'=0.002$ for the relevant WFIRST PSFs. The black $\times$ represents a nominal value of the IPC parameters in H4RG detectors.
 For each filter, the adaptive size in milliarcseconds (mas) and ellipticity $(e_1,e_2)$ without IPC is noted above the subplots.}
 \label{fig:aniso_e1}
\end{figure}
\begin{figure}
 \centering
 \includegraphics[width=\columnwidth]{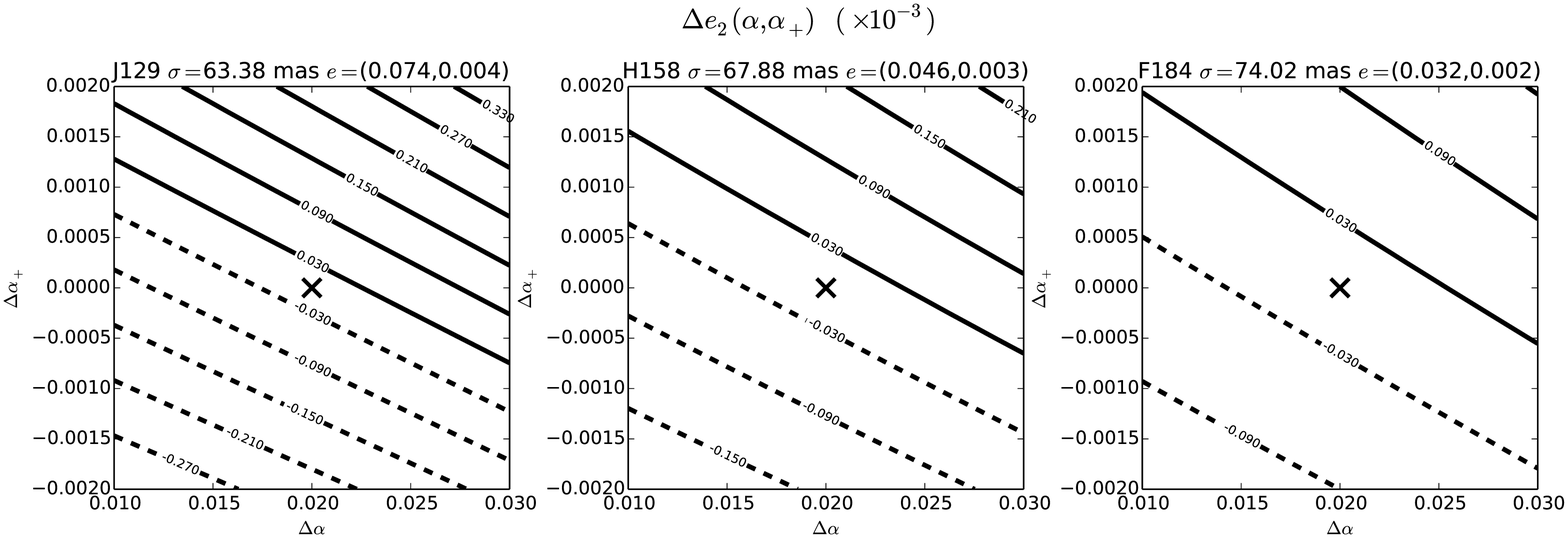}
 \caption{Contour plot of the change in the ellipticity component $e_2$ as a
   function of the $\alpha$ and $\alpha_+$ parameters in the IPC kernel in Eq.~\eqref{eq:threepK} with $\alpha'=0.002$ for the relevant WFIRST PSFs. The black $\times$ represents a nominal value of the IPC parameters in H4RG detectors.
 For each filter, the adaptive size in milliarcseconds (mas) and ellipticity $(e_1,e_2)$ without IPC is noted above the subplots.}
 \label{fig:aniso_e2}
\end{figure}

The second sub-class of IPC kernel allows for the anisotropy with a fixed value for the diagonal coupling $\alpha'$, set equal its nominal value of 0.002.
We consider this case since the PSF sizes and shapes are least sensitive to $\alpha'$.
Figs.~\ref{fig:aniso_e1}-~\ref{fig:aniso_e2} show the change to the components of PSF ellipticity.
The change in $e_2$ is about $\pm 5\% (\sim 10^{-4})$ over the entire range of $\alpha_+$ values we have considered whereas the change in $e_1$ is of the order $10^{-2}$, much bigger than the original $e_1$ value itself.
This is not surprising since both the anisotropy and the $e_1$ component are aligned along the axes. For a symmetric PSF with $e_2=0$, we expect the $e_2$ to remain 0 after the IPC.
The small change in $e_2$ is attributed to $e_2$ not being zero to begin with.
We verified that the relative increases in sizes have negligible dependence on the anisotropy (figure not shown).



\subsection{Error in PSF shapes from uncertainty in IPC parameters}
\label{subsec:errors}
If the IPC kernel is known precisely, then the effect can, in principle, be perfectly corrected.
However, errors in IPC parameters arise either because of measurement uncertainties or because of
the parameters varying (slowly) across the pixels.
In addition, there is a possibility of error due to adopting an incorrect IPC model, the use of a
$3\times 3$ kernel. Even with a perfect IPC correction scheme, errors in the IPC kernel that is
assumed and corrected for will propagate into
errors in PSF sizes and ellipticities. 
In this section, we investigate the magnitude of this effect, in all cases assuming a perfect
correction scheme.

\begin{figure}
 \centering
 \includegraphics[width=\columnwidth]{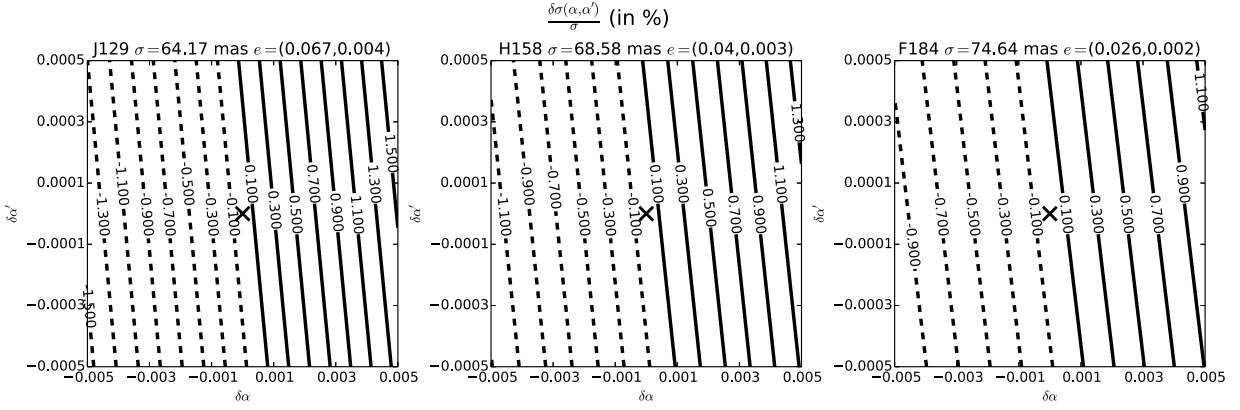}
 \caption{Contour plot of the relative change in the adaptive size (expressed as a percent) as a
   function of $\delta\alpha$ and $\delta\alpha'$, the deviation from their nominal values ($\alpha_0 = 0.02$ 
   and $\alpha'_0 = 0.002$) of the parameters in the IPC kernel in Eq.~\eqref{eq:twopK} for the relevant WFIRST PSFs.
   The black $\times$ corresponds to no deviation from the chosen nominal values of the IPC parameters in H4RG detectors.
 For each filter, the adaptive size in milliarcseconds (mas) and ellipticity $(e_1, e_2)$ without IPC is noted above the subplots.
}
 \label{fig:accu_nom_sigma}
\end{figure}

\begin{figure}
 \centering
  \includegraphics[width=\columnwidth]{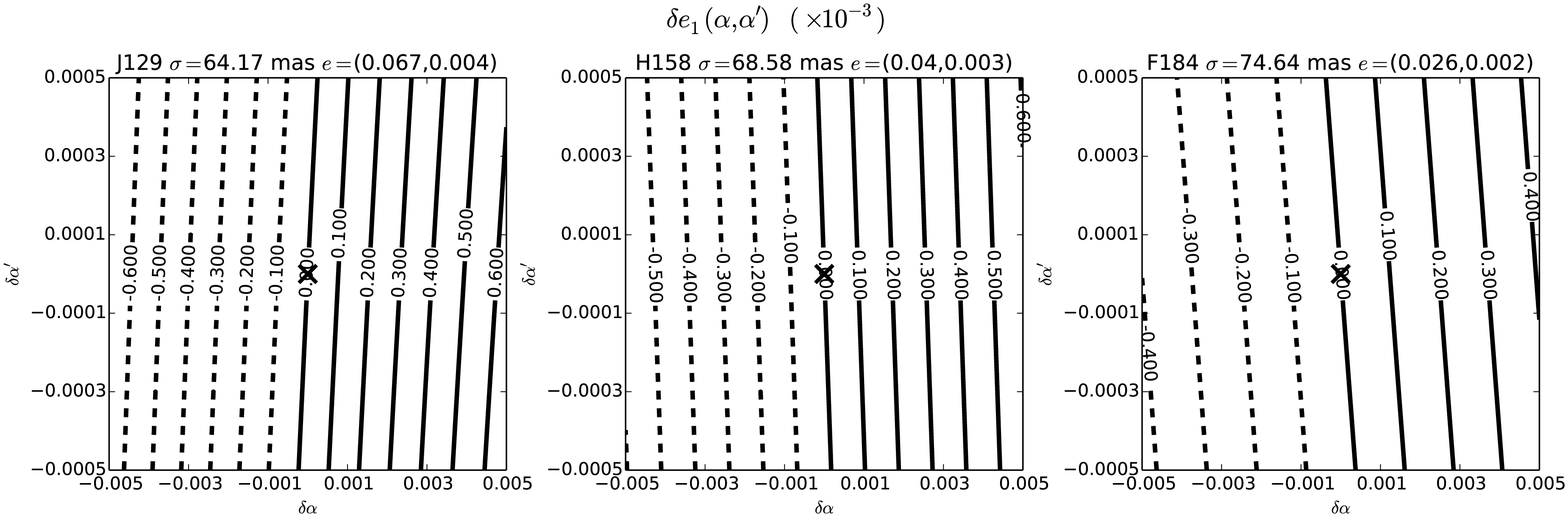}
  \caption{Contour plot of the absolute change (after multiplication by $10^3$) in the ellipticity component $e_1$ as a
   function of $\delta\alpha$ and $\delta\alpha'$, the deviation from their nominal values ($\alpha_0 = 0.02$ 
   and $\alpha'_0 = 0.002$) of the parameters in the IPC kernel in Eq.~\eqref{eq:twopK} for the relevant WFIRST PSFs.
   The black $\times$ corresponds to no deviation from the chosen nominal values of the IPC parameters in H4RG detectors.
 For each filter, the adaptive size in milliarcseconds (mas) and ellipticity $(e_1, e_2)$ without IPC is noted above the subplots.}
  \label{fig:accu_nom_e1}
\end{figure}

\begin{figure}
 \centering
 \includegraphics[width=\columnwidth]{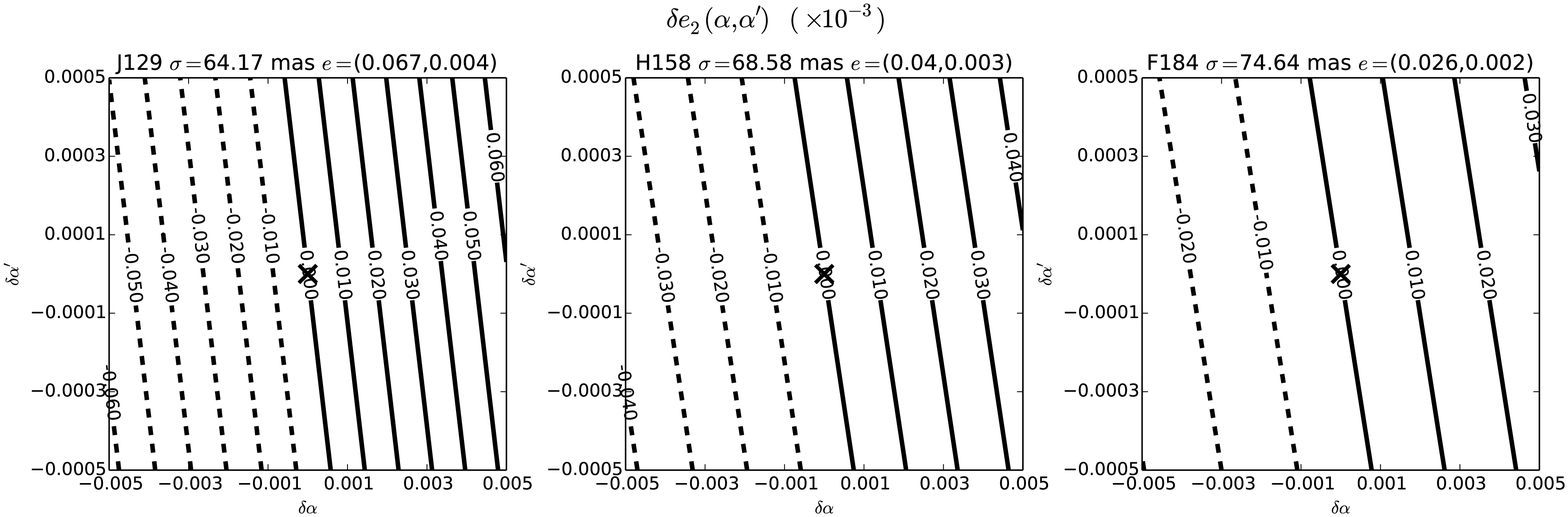}
  \caption{Contour plot of the absolute change (after multiplication by $10^3$) in the ellipticity component $e_2$ as a
   function of $\delta\alpha$ and $\delta\alpha'$, the deviation from their nominal values ($\alpha_0 = 0.02$ 
   and $\alpha'_0 = 0.002$) of the parameters in the IPC kernel in Eq.~\eqref{eq:twopK} for the relevant WFIRST PSFs.
   The black $\times$ corresponds to no deviation from the chosen nominal values of the IPC parameters in H4RG detectors.
 For each filter, the adaptive size in milliarcseconds (mas) and ellipticity $(e_1, e_2)$ without IPC is noted above the subplots.}
 \label{fig:accu_nom_e2}
\end{figure}

We start by considering the 2-parameter isotropic IPC kernel (Eq.~\ref{eq:twopK}) assuming that the
model is correct but its parameters have uncertainties. 
The issue at hand can be addressed by looking the change in the PSF sizes and shapes as a function of $\alpha$ and $\alpha'$, as done in Sec.~\ref{subsec:changes}; 
but now on a finer grid over the range of values of the IPC parameters where we expect them to lie.
In the ideal case of being able to correct for IPC exactly, the error in PSF size and shape will be due to $\delta\alpha$ and $\delta\alpha'$, the difference between the true
values of $\alpha$ and $\alpha'$ and their assumed values, denoted by $\alpha_0$ and $\alpha'_0$.
For nominal values of $\alpha_0=0.02$ and $\alpha'_0=0.002$, 
Figs.~\ref{fig:accu_nom_sigma}--\ref{fig:accu_nom_e2} show the error in PSF size and ellipticity as a function of errors in the parameters,
$\delta \alpha$ and $\delta \alpha'$. 
For a 10\% error in $\alpha$ and fixed $\alpha'$ ($\delta\alpha'=0$), the error in relative increase
in size is $\sim (5\pm0.5)$\%. 
All the contour lines in these three figures are
parallel and approximately equally spaced, which suggests that a linear fitting function could
describe these results very well.
Thus,
\begin{subequations}
 \begin{equation}
  \sigma'(\alpha,\alpha') - \sigma'(\alpha_0,\alpha'_0) \approx \delta\alpha \,R_\alpha(\alpha_0,\alpha'_0) + \delta\alpha'\,R_{\alpha'}(\alpha_0,\alpha'_0)
   \label{eq:linearized_sigma}
 \end{equation}
 \begin{equation}
  e'_k(\alpha,\alpha') - e'_k(\alpha_0,\alpha'_0) \approx \delta\alpha \,S_{k,\alpha}(\alpha_0,\alpha'_0) + \delta\alpha'\,S_{k,\alpha'}(\alpha_0,\alpha'_0), \;\;\;\text{for } k=1,2
 \end{equation}
\label{eq:linear_model}
\end{subequations}
for some set of coefficients $R_\alpha$, $R_{\alpha'}$, $S_{k,\alpha}$ and $S_{k,\alpha'}$ (for $k=1,2$), which are equivalent to appropriate partial derivatives evaluated
at the chosen nominal values of the parameters.


The coefficients in Eqs.~\eqref{eq:linear_model} were calculated by performing linear regression on an $11\times 11$ grid of $\delta\alpha$ and $\delta\alpha'$, with $\delta\alpha$ varying
uniformly from $-0.005$ to $+0.005$ and $\delta\alpha'$ varying uniformly from $-5\times10^{-4}$ to $+5\times10^{-4}$.
The values of the coefficients for the PSFs of the J129, H158 and F184 bandpasses are given in Table~\ref{table:isotropy_coeffs}.
We find that the quantities on the left hand side of Eqs.~\eqref{eq:linear_model} agree with the
linear model prediction with at most a 1\% error, for $\delta\alpha\approx 0.1\alpha_0$.  Thus, the errors in PSF sizes and ellipticities can simply be
specified by the six linear coefficients, $R_\alpha$, $R_{\alpha'}$, $S_{1,\alpha}$,
$S_{1,\alpha'}$, $S_{2,\alpha}$ and $S_{2,\alpha'}$, defined in Eq.~\ref{eq:linear_model}.

\begin{table}
 \centering
 \begin{tabular}{c|c|c|c|c|c|c}
  \hline
  Bandpass & $R_\alpha /\sigma(\alpha_0,\alpha_0')$ & $R_{\alpha'}/\sigma(\alpha_0,\alpha_0') $ & $S_{1,\alpha}$ & $S_{1,\alpha'}$ & $S_{2,\alpha}$ & $S_{2,\alpha'}$ \\
  \hline
  J129 & 2.9432 & 2.7691 & 0.1308& -0.063 & 0.0116 & 0.0134 \\
  H158 & 2.4457 & 2.5631 & 0.1171 & 0.0406 & 0.0076 & 0.0112 \\
  F184 & 2.0168 & 2.3741 & 0.0807 & 0.0604 & 0.0054 & 0.0084 \\
  \hline
 \end{tabular}
 \caption{List of the six coefficients given in Eqns.~\ref{eq:linear_model} for the J129, H158 and F184 bandpasses. The coefficients represent the `instantaneous rate' of change in PSF size
 and ellipticity with respect to the change in one of the IPC parameters about the nominal values for the case of isotropic kernel (Eq.~\ref{eq:twopK}).}
 \label{table:isotropy_coeffs}
\end{table}

We computed the coefficients corresponding to the PSFs with no supporting struts (not shown in the
figures or tables). 
When we do not include the struts (but do include obscuration and the expected level of
aberrations), $R_\alpha/\sigma(\alpha_0,\alpha_0')$ decreases by 10\% from its original
value when we include struts. This suggests that requirements on IPC may not depend very strongly on
the choice to include or exclude struts in the simulations.
The change in the $S$-type coefficients is comparable or even greater than the coefficients themselves by a factor of 2 in many cases
(and hence a change of sign in some cases),
indicating that the ellipticities are very sensitive to the diffraction spikes.

Instead of the adaptive size, if we consider the EE50 radius, the coefficient that is the analog of
$R_\alpha /\sigma(\alpha_0,\alpha'_0)$ is either comparable or slightly smaller (by 33\% or less) 
than the corresponding coefficient for the adaptive size. On the other hand, the coefficient that is
the analog of $R_{\alpha'} /
\sigma(\alpha_0,\alpha'_0)$ is consistently about 60\% larger than the corresponding one for the
adaptive size. Thus, the EE50 radius, like the unweighted moment (Sec.~\ref{sec:transformation}),
is more sensitive to the diagonal coupling than the adaptive size is.

\subsection{Effects on PSF due to anisotropy of IPC}

\begin{table}
\centering
\begin{tabular}{c|c|c|c|c|c|c}
\hline
Bandpass & $R_\alpha/\sigma(\alpha_0,\alpha_0')$ & $R_+/\sigma(\alpha_0,\alpha_0')$ & $S_{1,\alpha}$ & $S_{1,+}$ & $S_{2,\alpha}$ & $S_{2,+}$ \\\hline
J129 & 2.9429 & 0.0383 & 0.1308 & 4.1516 & 0.0116 & 0.1165 \\
H158 & 2.4454 & 0.0435 & 0.1171 & 3.7389 & 0.0075 & 0.0704 \\
F184 & 2.0166 & 0.0336 & 0.0807 & 3.2975 & 0.0054 & 0.045 \\
\hline
\end{tabular}
 \caption{List of the six coefficients given in Eqns.~\ref{eq:anisotropic_linear_model} for the J129, H158 and F184 bandpasses. The coefficients represent the `instantaneous rate' of change in PSF size
 and ellipticity with respect to the change in one of the nearest neighbor coupling parameters about the nominal values for the case of anisotropic kernel (Eq.~\ref{eq:threepK}).}
\label{table:anisotropy_coefficients}
\end{table}

We now address the effect of a slight anisotropy that is commonly observed in the IPC kernel on the PSF shapes.
The difference between the nearest neighbor coupling along the axes of H4RG detectors is expected to be small, as in for WFC3 (Eq.~\ref{eq:KWFC3}).
This lets us handle the anisotropy perturbatively as we handled the uncertainties in the coupling in Sec.~\ref{subsec:errors}.
If we denote the average nearest neighbor coupling along both the directions as $\alpha$, we consider the exact
coupling along $x$ and $y$ axes as a small deviation from $\alpha$, which we denote as $\alpha_+$. Generalizing the Eqns.~\ref{eq:linear_model},
we now write
\begin{subequations}
\begin{equation}
 \delta\sigma' = \delta\alpha\,R_\alpha + \delta\alpha_+\,R_+ ,
\end{equation}
\begin{equation}
 \delta e'_k = \delta\alpha\,S_{k,\alpha} + \delta\alpha_+\,S_{k,+} , \;\;\; \text{for } k=1,2. 
\end{equation}
\label{eq:anisotropic_linear_model}
\end{subequations}

Naively, we expect the anisotropy to have only a weak effect on the PSF sizes and a relatively stronger effect on the ellipticity, compared to that from the isotropic term $\alpha$.
We verify this and quantify the effect from simulations.
In our simulations to compute the coefficients in Eq.~\ref{eq:anisotropic_linear_model}, we varied $\alpha_x$ defined as $\alpha+\alpha_+$ and $\alpha_y$ defined as $\alpha-\alpha_+$ uniformly on a $11\times 11$ grid from $0.02-0.005$ to $0.02+0.005$.
We held $\alpha'$ constant at its nominal value of $2\times 10^{-3}$ for simplicity. 
The best fit parameters for $R_\alpha$, $R_+$, $S_{k,\alpha}$ and $S_{k,+}$ in Eqns.~\ref{eq:anisotropic_linear_model} 
are found from linear regression. If the model is valid,
then we should expect the $R_\alpha$ values to agree with their earlier values. Table~\ref{table:anisotropy_coefficients} lists these coefficients for the J129, H158 and F184 bandpasses
and indeed the $R_\alpha$ values agree for all of them. 

\section{IPC in the context of WFIRST requirements}
It is informative to place the results of our analysis in Sec.~\ref{sec:results} into context, comparing IPC-related shape measurement
errors to the requirements for the WFIRST PSF as outlined in ~\cite{spergel2015wide}.
We highlight again that the PSFs used in our analysis do not include jitter and charge diffusion. 
These effects will tend to increase the size of the PSFs and somewhat reduce the relative effects of IPC. 
Thus, not including these effects would tend to push our bounds on the IPC parameters to the slightly pessimistic/conservative side.

While the maximum allowed PSF sizes are given in terms of the half-light radii, the maximum tolerable error on PSF sizes are given in terms of the second moments.
Since the adaptive moments are sensitive mainly to the core of the PSF, our overall conclusions based on adaptive moments are fairly independent of the choice of
including the diffraction spikes. This can be seen from Table~\ref{tab:size_hlr}, where the adaptive sizes change by $\lesssim 1.6\%$ whether or not we consider the diffraction
spikes as a part of the PSF.

Let us suppose that it is possible to measure the IPC parameters to within errors of $\delta\alpha=\delta\alpha'=\delta\alpha_+=10^{-4}$.  These could be simple calibration errors or perhaps due to some spatial or temporal IPC variation that went uncharacterized.  If we model the WFIRST PSF with these parameters, we see that the worst PSF shape errors are in the J band, with $\delta\sigma /\sigma\approx5.75\times 10^{-4}$
(and from Eq.~\ref{eq:tr_sigma}, $\delta{\text{tr}({\bf M})}/{\text{tr}({\bf M})} \lesssim 1.15\times 10^{-3}$) and $\delta e_1 \approx 4.35\times 10^{-4}$. These values
are obtained from Eqs.~\ref{eq:linear_model} and ~\ref{eq:anisotropic_linear_model}, with values for the coefficients taken from Tables~\ref{table:isotropy_coeffs} and~\ref{table:anisotropy_coefficients}.
For the High Latitude Imaging Survey, the tolerance on the relative error of trace of the second moments of the PSFs is set as 0.093\%, and the required knowledge of PSF ellipticity is $4.7\times 10^{-4}$ per component~\citep{spergel2015wide}. 
These tolerances might evolve as the details of the survey are finalized. Thus, our model would have errors comparable to the tolerances for the entire survey.  In practice, a PSF model will be fit to on-sky measurements, and shape errors caused by a misestimate of IPC parameters will be absorbed by other parameters in the PSF model.  This freedom relaxes requirements for the IPC parameters.  To set definitive IPC requirements for WFIRST, we need to consider the planned shape measurement strategy and other sources of shape measurement error, which are still being studied.  However, we have shown that the effect of IPC on shape measurement is sufficiently large that our ability to calibrate it out should not be taken for granted.

\section{Conclusion and future work}\label{sec:conclusion}
We have explored the impact of interpixel capacitance, an effect that will be relevant for surveys
such as WFIRST that will use near-infrared detectors, on the point-spread function, including its
size and shape.   To carry out this work, we have created a new WFIRST module in the
publicly-available GalSim software package.  
Using this software, we have determined linear fitting formulae that describe how 
the PSF size and shape change with the level of IPC, including the effects of changing different
parts of a two-parameter IPC model.  
Our simulations show that the level of IPC that is present in state-of-the-art detector technology
will increase the typical PSF
sizes by $\sim 5$\% for WFIRST. This is roughly the case irrespective of whether we include the supporting struts in simulating the PSFs or not.

The isotropic IPC kernel also changes the ellipticity of the PSFs. The changes in $e_1$ and $e_2$
are an order of magnitude smaller than the expected PSF $e_1$ and $e_2$ themselves. 
These results may be useful inputs into WFIRST hardware requirements.

There are schemes~\citep{mccullough2008correction} for removing the effect of IPC from astronomical images.  However, these schemes
rely on knowledge of the IPC.  Thus, we also consider the scenario where the IPC is assumed to be at the nominally
expected level, but there are actually systematic deviations from that level, and derive linear fitting
formulae for errors on PSF sizes and shapes in the different WFIRST passbands. 
Assuming a perfect IPC correction scheme, a 10\% error in determining the nearest neighbor coupling
results in relative errors that are of the order $5\times 10^{-3}$ in the PSF sizes.
For the errors in PSF sizes and shapes to be within the allowed limits for WFIRST, the uncertainty in the parameters $\alpha$, $\alpha'$ and $\alpha_+$ must be much smaller than $10^{-4}$.

Our results represent an important step towards placing requirements on hardware that affects the
PSF for the WFIRST survey.  Future work will use these to consider their impact on the scientific
measurements of interest, such as weak lensing.
The modifications of the PSFs due to imperfect knowledge of IPC can contribute towards shear calibration biases
when carrying out the process of PSF correction to estimate weak lensing shears (Kannawadi et.\ al.,
in prep). Even if we had perfect knowledge of IPC, the correlation in the noise caused due to IPC will affect the shear calibration biases.
We analyze these effects of IPC in the presence of noise in a future work (Kannawadi et.\ al., in prep).

Another interesting direction for future work would be to consider the interplay between different detector effects, such as IPC and read noise.
Since the introduction of read noise happens at a later stage compared to IPC, any IPC correction
scheme would correlate the read noise. We defer exploration of the impact of correlated read
noise to future work.

\subsection*{Acknowledgement}
The authors thank Roger Smith, Bernard Rauscher and Andr\'{e}s Plazas Malag\'{o}n for many useful discussions
and Mike Jarvis and Joshua Meyers for their inputs in developing the GalSim WFIRST module. We thank Edward Cheng of Conceptual Analytics for his comments in improving the manuscript
and the referee, David Spergel, for correcting a few minor errors in the original version of the manuscript.
This work was carried out in part at the Jet Propulsion Laboratory (JPL), a NASA center run by California Institute of Technology.
The authors acknowledge funding from WFIRST study office. 
C.M.H. is supported by the US Department of Energy, the Packard Foundation, and the Simons Foundation.

\bibliographystyle{authordate1}
\bibliography{ref}

\begin{thebibliography}{}

\bibitem[\protect\citename{{Bartelmann} \& {Schneider},
  }2001]{2001PhR...340..291B}
{Bartelmann}, M., \& {Schneider}, P. 2001.
\newblock {Weak gravitational lensing}.
\newblock {\em \physrep}, {\bf 340}(Jan.), 291--472.

\bibitem[\protect\citename{Bebek {\em et~al.\ }\relax,
  }2015]{1748-0221-10-05-C05026}
Bebek, C.J., Emes, J.H., Groom, D.E., Haque, S., Holland, S.E., Karcher, A.,
  Kolbe, W.F., Lee, J.S., Palaio, N.P., \& Wang, G. 2015.
\newblock {CCD development for the Dark Energy Spectroscopic Instrument}.
\newblock {\em Journal of Instrumentation}, {\bf 10}(05), C05026.

\bibitem[\protect\citename{{Becker} {\em et~al.\ }\relax,
  }2015]{2015arXiv150705598B}
{Becker}, M.~R., {Troxel}, M.~A., {MacCrann}, N., {Krause}, E., {Eifler},
  T.~F., {Friedrich}, O., {Nicola}, A., {Refregier}, A., {Amara}, A., {Bacon},
  D., {Bernstein}, G.~M., {Bonnett}, C., {Bridle}, S.~L., {Busha}, M.~T.,
  {Chang}, C., {Dodelson}, S., {Erickson}, B., {Evrard}, A.~E., {Frieman}, J.,
  {Gaztanaga}, E., {Gruen}, D., {Hartley}, W., {Jain}, B., {Jarvis}, M.,
  {Kacprzak}, T., {Kirk}, D., {Kravtsov}, A., {Leistedt}, B., {Rykoff}, E.~S.,
  {Sabiu}, C., {Sanchez}, C., {Seo}, H., {Sheldon}, E., {Wechsler}, R.~H.,
  {Zuntz}, J., {Abbott}, T., {Abdalla}, F.~B., {Allam}, S., {Armstrong}, R.,
  {Banerji}, M., {Bauer}, A.~H., {Benoit-Levy}, A., {Bertin}, E., {Brooks}, D.,
  {Buckley-Geer}, E., {Burke}, D.~L., {Capozzi}, D., {Carnero Rosell}, A.,
  {Carrasco Kind}, M., {Carretero}, J., {Castander}, F.~J., {Crocce}, M.,
  {Cunha}, C.~E., {D'Andrea}, C.~B., {da Costa}, L.~N., {DePoy}, D.~L.,
  {Desai}, S., {Diehl}, H.~T., {Dietrich}, J.~P., {Doel}, P., {Fausti Neto},
  A., {Fernandez}, E., {Finley}, D.~A., {Flaugher}, B., {Fosalba}, P.,
  {Gerdes}, D.~W., {Gruendl}, R.~A., {Gutierrez}, G., {Honscheid}, K., {James},
  D.~J., {Kuehn}, K., {Kuropatkin}, N., {Lahav}, O., {Li}, T.~S., {Lima}, M.,
  {Maia}, M.~A.~G., {March}, M., {Martini}, P., {Melchior}, P., {Miller},
  C.~J., {Miquel}, R., {Mohr}, J.~J., {Nichol}, R.~C., {Nord}, B., {Ogando},
  R., {Plazas}, A.~A., {Reil}, K., {Romer}, A.~K., {Roodman}, A., {Sako}, M.,
  {Sanchez}, E., {Scarpine}, V., {Schubnell}, M., {Sevilla-Noarbe}, I.,
  {Smith}, R.~C., {Soares-Santos}, M., {Sobreira}, F., {Suchyta}, E.,
  {Swanson}, M.~E.~C., /IPC~{Tarle}, G., {Thaler}, J., {Thomas}, D., {Vikram},
  V., {Walker}, A.~R., \& {The DES Collaboration}. 2015.
\newblock {Cosmic Shear Measurements with DES Science Verification Data}.
\newblock {\em ArXiv e-prints}, July.

\bibitem[\protect\citename{{Bernstein} \& {Jarvis}, }2002]{BJ02}
{Bernstein}, G.~M., \& {Jarvis}, M. 2002.
\newblock {Shapes and Shears, Stars and Smears: Optimal Measurements for Weak
  Lensing}.
\newblock {\em \aj}, {\bf 123}(Feb.), 583--618.

\bibitem[\protect\citename{Biesiadzinski {\em et~al.\ }\relax,
  }2011]{2011reciprocity}
Biesiadzinski, T, Lorenzon, W, Newman, R, Schubnell, M, Tarl{\'e}, G, \&
  Weaverdyck, C. 2011.
\newblock Reciprocity Failure in HgCdTe Detectors: Measurements and Mitigation.
\newblock {\em Publications of the Astronomical Society of the Pacific}, {\bf
  123}(906), 958--963.

\bibitem[\protect\citename{{Coupon} {\em et~al.\ }\relax,
  }2015]{2015MNRAS.449.1352C}
{Coupon}, J., {Arnouts}, S., {van Waerbeke}, L., {Moutard}, T., {Ilbert}, O.,
  {van Uitert}, E., {Erben}, T., {Garilli}, B., {Guzzo}, L., {Heymans}, C.,
  {Hildebrandt}, H., {Hoekstra}, H., {Kilbinger}, M., {Kitching}, T.,
  {Mellier}, Y., {Miller}, L., {Scodeggio}, M., {Bonnett}, C., {Branchini}, E.,
  {Davidzon}, I., {De Lucia}, G., {Fritz}, A., {Fu}, L., {Hudelot}, P.,
  {Hudson}, M.~J., {Kuijken}, K., {Leauthaud}, A., {Le F{\`e}vre}, O.,
  {McCracken}, H.~J., {Moscardini}, L., {Rowe}, B.~T.~P., {Schrabback}, T.,
  {Semboloni}, E., \& {Velander}, M. 2015.
\newblock {The galaxy-halo connection from a joint lensing, clustering and
  abundance analysis in the CFHTLenS/VIPERS field}.
\newblock {\em \mnras}, {\bf 449}(May), 1352--1379.

\bibitem[\protect\citename{Crouzet {\em et~al.\ }\relax,
  }2012]{doi:10.1117/12.924968}
Crouzet, P-E., ter Haar, J., de~Wit, F., Beaufort, T., Butler, B., Smit, H.,
  van~der Luijt, C., \& Martin, D. 2012.
\newblock Characterization of HAWAII-2RG detector and SIDECAR ASIC for the
  Euclid mission at ESA.
\newblock {\em Proc. SPIE}, {\bf 8453}, 84531R--84531R--15.

\bibitem[\protect\citename{Estrada {\em et~al.\ }\relax,
  }2006]{doi:10.1117/12.672596}
Estrada, J., Abbott, T., Angstadt, B., Buckley-Geer, L., Brown, M., Campa, J.,
  Cardiel, L., Cease, H., Flaugher, B., Dawson, K., Derylo, G., Diehl, H.~T.,
  Gruenendahl, S., Karliner, I., Merrit, W., Moore, P., Moore, T.~C., Roe, N.,
  Scarpine, V., Schmidt, R., Schubnel, M., Shaw, T., Stuermer, W., \& Thaler,
  J. 2006.
\newblock CCD testing and characterization for dark energy survey.
\newblock {\em Proc. SPIE}, {\bf 6269}, 62693K--62693K--15.

\bibitem[\protect\citename{Fox {\em et~al.\ }\relax,
  }2009]{2008SPIE.7021E..23F}
Fox, Ori, Waczynski, Augustyn, Wen, Yiting, Foltz, Roger~D, Hill, Robert~J,
  Kimble, Randy~A, Malumuth, Eliot, \& Rauscher, Bernard~J. 2009.
\newblock {The 55Fe X-Ray Energy Response of Mercury Cadmium Telluride
  Near-Infrared Detector Arrays}.
\newblock {\em Publications of the Astronomical Society of the Pacific}, {\bf
  121}(881), 743.

\bibitem[\protect\citename{{Fruchter}, }2011]{iDrizzle}
{Fruchter}, A.~S. 2011.
\newblock {A New Method for Band-limited Imaging with Undersampled Detectors}.
\newblock {\em \pasp}, {\bf 123}(Apr.), 497--502.

\bibitem[\protect\citename{{Fruchter} \& {Hook}, }2002]{Drizzle}
{Fruchter}, A.~S., \& {Hook}, R.~N. 2002.
\newblock {Drizzle: A Method for the Linear Reconstruction of Undersampled
  Images}.
\newblock {\em \pasp}, {\bf 114}(Feb.), 144--152.

\bibitem[\protect\citename{{Gardner} {\em et~al.\ }\relax, }2006]{JWST}
{Gardner}, J.~P., {Mather}, J.~C., {Clampin}, M., {Doyon}, R., {Greenhouse},
  M.~A., {Hammel}, H.~B., {Hutchings}, J.~B., {Jakobsen}, P., {Lilly}, S.~J.,
  {Long}, K.~S., {Lunine}, J.~I., {McCaughrean}, M.~J., {Mountain}, M.,
  {Nella}, J., {Rieke}, G.~H., {Rieke}, M.~J., {Rix}, H.-W., {Smith}, E.~P.,
  {Sonneborn}, G., {Stiavelli}, M., {Stockman}, H.~S., {Windhorst}, R.~A., \&
  {Wright}, G.~S. 2006.
\newblock {The James Webb Space Telescope}.
\newblock {\em \ssr}, {\bf 123}(Apr.), 485--606.

\bibitem[\protect\citename{{Green} {\em et~al.\ }\relax, }2012]{WFIRST_final}
{Green}, J., {Schechter}, P., {Baltay}, C., {Bean}, R., {Bennett}, D., {Brown},
  R., {Conselice}, C., {Donahue}, M., {Fan}, X., {Gaudi}, B.~S., {Hirata}, C.,
  {Kalirai}, J., {Lauer}, T., {Nichol}, B., {Padmanabhan}, N., {Perlmutter},
  S., {Rauscher}, B., {Rhodes}, J., {Roellig}, T., {Stern}, D., {Sumi}, T.,
  {Tanner}, A., {Wang}, Y., {Weinberg}, D., {Wright}, E., {Gehrels}, N.,
  {Sambruna}, R., {Traub}, W., {Anderson}, J., {Cook}, K., {Garnavich}, P.,
  {Hillenbrand}, L., {Ivezic}, Z., {Kerins}, E., {Lunine}, J., {McDonald}, P.,
  {Penny}, M., {Phillips}, M., {Rieke}, G., {Riess}, A., {van der Marel}, R.,
  {Barry}, R.~K., {Cheng}, E., {Content}, D., {Cutri}, R., {Goullioud}, R.,
  {Grady}, K., {Helou}, G., {Jackson}, C., {Kruk}, J., {Melton}, M., {Peddie},
  C., {Rioux}, N., \& {Seiffert}, M. 2012.
\newblock {Wide-Field InfraRed Survey Telescope (WFIRST) Final Report}.
\newblock {\em ArXiv e-prints}, Aug.

\bibitem[\protect\citename{{Han} {\em et~al.\ }\relax,
  }2015]{2015MNRAS.446.1356H}
{Han}, J., {Eke}, V.~R., {Frenk}, C.~S., {Mandelbaum}, R., {Norberg}, P.,
  {Schneider}, M.~D., {Peacock}, J.~A., {Jing}, Y., {Baldry}, I.,
  {Bland-Hawthorn}, J., {Brough}, S., {Brown}, M.~J.~I., {Liske}, J.,
  {Loveday}, J., \& {Robotham}, A.~S.~G. 2015.
\newblock {Galaxy And Mass Assembly (GAMA): the halo mass of galaxy groups from
  maximum-likelihood weak lensing}.
\newblock {\em \mnras}, {\bf 446}(Jan.), 1356--1379.

\bibitem[\protect\citename{{Heymans} {\em et~al.\ }\relax,
  }2013]{2013MNRAS.432.2433H}
{Heymans}, C., {Grocutt}, E., {Heavens}, A., {Kilbinger}, M., {Kitching},
  T.~D., {Simpson}, F., {Benjamin}, J., {Erben}, T., {Hildebrandt}, H.,
  {Hoekstra}, H., \& et~al. 2013.
\newblock {CFHTLenS tomographic weak lensing cosmological parameter
  constraints: Mitigating the impact of intrinsic galaxy alignments}.
\newblock {\em \mnras}, {\bf 432}(July), 2433--2453.

\bibitem[\protect\citename{Hilbert \& McCullough, }2011]{hilbert2011interpixel}
Hilbert, B, \& McCullough, P. 2011.
\newblock Interpixel Capacitance in the IR Channel: Measurements Made On Orbit.

\bibitem[\protect\citename{Hirata \& Seljak, }2003]{HS03}
Hirata, Christopher, \& Seljak, Uroš. 2003.
\newblock Shear calibration biases in weak-lensing surveys.
\newblock {\em Monthly Notices of the Royal Astronomical Society}, {\bf
  343}(2), 459--480.

\bibitem[\protect\citename{{Hoekstra} \& {Jain}, }2008]{2008ARNPS..58...99H}
{Hoekstra}, H., \& {Jain}, B. 2008.
\newblock {Weak Gravitational Lensing and Its Cosmological Applications}.
\newblock {\em Annual Review of Nuclear and Particle Science}, {\bf 58}(Nov.),
  99--123.

\bibitem[\protect\citename{{Hudson} {\em et~al.\ }\relax,
  }2015]{2015MNRAS.447..298H}
{Hudson}, M.~J., {Gillis}, B.~R., {Coupon}, J., {Hildebrandt}, H., {Erben}, T.,
  {Heymans}, C., {Hoekstra}, H., {Kitching}, T.~D., {Mellier}, Y., {Miller},
  L., {Van Waerbeke}, L., {Bonnett}, C., {Fu}, L., {Kuijken}, K., {Rowe}, B.,
  {Schrabback}, T., {Semboloni}, E., {van Uitert}, E., \& {Velander}, M. 2015.
\newblock {CFHTLenS: co-evolution of galaxies and their dark matter haloes}.
\newblock {\em \mnras}, {\bf 447}(Feb.), 298--314.

\bibitem[\protect\citename{{Jee} {\em et~al.\ }\relax,
  }2013]{2013ApJ...765...74J}
{Jee}, M.~J., {Tyson}, J.~A., {Schneider}, M.~D., {Wittman}, D., {Schmidt}, S.,
  \& {Hilbert}, S. 2013.
\newblock {Cosmic Shear Results from the Deep Lens Survey. I. Joint Constraints
  on {$\Omega$}$_{ M }$ and {$\sigma$}$_{8}$ with a Two-dimensional Analysis}.
\newblock {\em \apj}, {\bf 765}(Mar.), 74.

\bibitem[\protect\citename{{Kaiser} {\em et~al.\ }\relax, }1995]{KSB95}
{Kaiser}, N., {Squires}, G., \& {Broadhurst}, T. 1995.
\newblock {A Method for Weak Lensing Observations}.
\newblock {\em \apj}, {\bf 449}(Aug.), 460.

\bibitem[\protect\citename{Kimble {\em et~al.\ }\relax, }2008]{WFC3}
Kimble, Randy~A., MacKenty, John~W., O'Connell, Robert~W., \& Townsend,
  Jacqueline~A. 2008.
\newblock Wide Field Camera 3: a powerful new imager for the Hubble Space
  Telescope.
\newblock {\em Proc. SPIE}, {\bf 7010}, 70101E--70101E--12.

\bibitem[\protect\citename{{Lauer}, }1999]{1999Lauer}
{Lauer}, T.~R. 1999.
\newblock {Combining Undersampled Dithered Images}.
\newblock {\em \pasp}, {\bf 111}(Feb.), 227--237.

\bibitem[\protect\citename{{Leauthaud} {\em et~al.\ }\relax,
  }2012]{2012ApJ...744..159L}
{Leauthaud}, A., {Tinker}, J., {Bundy}, K., {Behroozi}, P.~S., {Massey}, R.,
  {Rhodes}, J., {George}, M.~R., {Kneib}, J.-P., {Benson}, A., {Wechsler},
  R.~H., {Busha}, M.~T., {Capak}, P., {Cort{\^e}s}, M., {Ilbert}, O.,
  {Koekemoer}, A.~M., {Le F{\`e}vre}, O., {Lilly}, S., {McCracken}, H.~J.,
  {Salvato}, M., {Schrabback}, T., {Scoville}, N., {Smith}, T., \& {Taylor},
  J.~E. 2012.
\newblock {New Constraints on the Evolution of the Stellar-to-dark Matter
  Connection: A Combined Analysis of Galaxy-Galaxy Lensing, Clustering, and
  Stellar Mass Functions from z = 0.2 to z =1}.
\newblock {\em \apj}, {\bf 744}(Jan.), 159.

\bibitem[\protect\citename{{Mandelbaum} {\em et~al.\ }\relax,
  }2013]{2013MNRAS.432.1544M}
{Mandelbaum}, R., {Slosar}, A., {Baldauf}, T., {Seljak}, U., {Hirata}, C.~M.,
  {Nakajima}, R., {Reyes}, R., \& {Smith}, R.~E. 2013.
\newblock {Cosmological parameter constraints from galaxy-galaxy lensing and
  galaxy clustering with the SDSS DR7}.
\newblock {\em \mnras}, {\bf 432}(June), 1544--1575.

\bibitem[\protect\citename{{Mandelbaum} {\em et~al.\ }\relax,
  }2015]{GREAT3_results1}
{Mandelbaum}, R., {Rowe}, B., {Armstrong}, R., {Bard}, D., {Bertin}, E.,
  {Bosch}, J., {Boutigny}, D., {Courbin}, F., {Dawson}, W.~A., {Donnarumma},
  A., {Fenech Conti}, I., {Gavazzi}, R., {Gentile}, M., {Gill}, M.~S.~S.,
  {Hogg}, D.~W., {Huff}, E.~M., {Jee}, M.~J., {Kacprzak}, T., {Kilbinger}, M.,
  {Kuntzer}, T., {Lang}, D., {Luo}, W., {March}, M.~C., {Marshall}, P.~J.,
  {Meyers}, J.~E., {Miller}, L., {Miyatake}, H., {Nakajima}, R., {Ngol{\'e}
  Mboula}, F.~M., {Nurbaeva}, G., {Okura}, Y., {Paulin-Henriksson}, S.,
  {Rhodes}, J., {Schneider}, M.~D., {Shan}, H., {Sheldon}, E.~S., {Simet}, M.,
  {Starck}, J.-L., {Sureau}, F., {Tewes}, M., {Zarb Adami}, K., {Zhang}, J., \&
  {Zuntz}, J. 2015.
\newblock {GREAT3 results - I. Systematic errors in shear estimation and the
  impact of real galaxy morphology}.
\newblock {\em \mnras}, {\bf 450}(July), 2963--3007.

\bibitem[\protect\citename{McCullough, }2008]{mccullough2008correction}
McCullough, P. 2008.
\newblock Inter-pixel capacitance: prospects for deconvolution.
\newblock {\em Instrument Science Report WFC3}, {\bf 26}.

\bibitem[\protect\citename{McCullough {\em et~al.\ }\relax,
  }2007]{mccullough2007measurement}
McCullough, Peter~R, Regan, M, Bergeron, L, \& Lindsay, K. 2007.
\newblock Measurement of the Quantum efficiency of an HgCdTe Infrared sensor
  Array.
\newblock {\em BULLETIN-AMERICAN ASTRONOMICAL SOCIETY}, {\bf 39}(1), 086.

\bibitem[\protect\citename{Moore {\em et~al.\ }\relax,
  }2004]{moore2004interpixel}
Moore, Andrew~C, Ninkov, Zoran, \& Forrest, William~J. 2004.
\newblock Interpixel capacitance in nondestructive focal plane arrays.
\newblock {\em Pages  204--215 of:} {\em Optical Science and Technology, SPIE's
  48th Annual Meeting}.
\newblock International Society for Optics and Photonics.

\bibitem[\protect\citename{Moore {\em et~al.\ }\relax, }2006]{moore2006quantum}
Moore, Andrew~C, Ninkov, Zoran, \& Forrest, William~J. 2006.
\newblock Quantum efficiency overestimation and deterministic cross talk
  resulting from interpixel capacitance.
\newblock {\em Optical Engineering}, {\bf 45}(7), 076402--076402.

\bibitem[\protect\citename{Noll, }1976]{Noll76}
Noll, Robert~J. 1976.
\newblock Zernike polynomials and atmospheric turbulence.
\newblock {\em J. Opt. Soc. Am.}, {\bf 66}(3), 207--211.

\bibitem[\protect\citename{O'Connor, }2015]{1748-0221-10-05-C05010}
O'Connor, P. 2015.
\newblock {Crosstalk in multi-output CCDs for LSST}.
\newblock {\em Journal of Instrumentation}, {\bf 10}(05), C05010.

\bibitem[\protect\citename{Okura {\em et~al.\ }\relax,
  }2015]{1748-0221-10-08-C08010}
Okura, Y., Plazas, A.A., May, M., \& Tamagawa, T. 2015.
\newblock {Spurious shear induced by the tree rings of the LSST CCDs}.
\newblock {\em Journal of Instrumentation}, {\bf 10}(08), C08010.

\bibitem[\protect\citename{Pasquale {\em et~al.\ }\relax,
  }2014]{pasquale2014optical}
Pasquale, Bert, Content, David, Kruk, Jeffery, Vaughnn, David, Gong, Qian,
  Howard, Joseph, Jurling, Alden, Seals, Len, Mentzell, Eric, Armani, Nerses,
  \& Kuan, Gary. 2014.
\newblock Optical design of the WFIRST-AFTA wide-field instrument.
\newblock {\em Proc. SPIE}, {\bf 9293}, 929305--929305--8.

\bibitem[\protect\citename{{Pullen} {\em et~al.\ }\relax,
  }2015]{2015MNRAS.449.4326P}
{Pullen}, A.~R., {Alam}, S., \& {Ho}, S. 2015.
\newblock {Probing gravity at large scales through CMB lensing}.
\newblock {\em \mnras}, {\bf 449}(June), 4326--4335.

\bibitem[\protect\citename{{Refregier}, }2003]{2003ARA&A..41..645R}
{Refregier}, A. 2003.
\newblock {Weak Gravitational Lensing by Large-Scale Structure}.
\newblock {\em \araa}, {\bf 41}, 645--668.

\bibitem[\protect\citename{{Reyes} {\em et~al.\ }\relax,
  }2010]{2010Natur.464..256R}
{Reyes}, R., {Mandelbaum}, R., {Seljak}, U., {Baldauf}, T., {Gunn}, J.~E.,
  {Lombriser}, L., \& {Smith}, R.~E. 2010.
\newblock {Confirmation of general relativity on large scales from weak lensing
  and galaxy velocities}.
\newblock {\em \nat}, {\bf 464}(Mar.), 256--258.

\bibitem[\protect\citename{{Rowe} {\em et~al.\ }\relax, }2011]{IMCOM_algo}
{Rowe}, B., {Hirata}, C., \& {Rhodes}, J. 2011.
\newblock {Optimal Linear Image Combination}.
\newblock {\em \apj}, {\bf 741}(Nov.), 46.

\bibitem[\protect\citename{{Rowe} {\em et~al.\ }\relax, }2015]{GalSim}
{Rowe}, B.~T.~P., {Jarvis}, M., {Mandelbaum}, R., {Bernstein}, G.~M., {Bosch},
  J., {Simet}, M., {Meyers}, J.~E., {Kacprzak}, T., {Nakajima}, R., {Zuntz},
  J., {Miyatake}, H., {Dietrich}, J.~P., {Armstrong}, R., {Melchior}, P., \&
  {Gill}, M.~S.~S. 2015.
\newblock {GALSIM: The modular galaxy image simulation toolkit}.
\newblock {\em Astronomy and Computing}, {\bf 10}(Apr.), 121--150.

\bibitem[\protect\citename{{Schneider}, }2006]{2006glsw.conf..269S}
{Schneider}, P. 2006.
\newblock {Part 3: Weak gravitational lensing}.
\newblock {\em Pages  269--451 of:} {Meylan}, G., {Jetzer}, P., {North}, P.,
  {Schneider}, P., {Kochanek}, C.~S., \& {Wambsganss}, J. (eds), {\em Saas-Fee
  Advanced Course 33: Gravitational Lensing: Strong, Weak and Micro}.
\newblock Springer-Verlag Berlin Heidelberg.

\bibitem[\protect\citename{{Seshadri} {\em et~al.\ }\relax,
  }2008]{Seshadri2008mapping}
{Seshadri}, S., Cole, D.~M., Hancock, B.~R., \& Smith, R.~M. 2008.
\newblock Mapping electrical crosstalk in pixelated sensor arrays.
\newblock {\em Proc. SPIE}, {\bf 7021}, 702104--702104--11.

\bibitem[\protect\citename{{Shapiro} {\em et~al.\ }\relax,
  }2013]{IMCOM_WLsystematics}
{Shapiro}, C., {Rowe}, B.~T.~P., {Goodsall}, T., {Hirata}, C., {Fucik}, J.,
  {Rhodes}, J., {Seshadri}, S., \& {Smith}, R. 2013.
\newblock {Weak Gravitational Lensing Systematics from Image Combination}.
\newblock {\em \pasp}, {\bf 125}(Dec.), 1496--1513.

\bibitem[\protect\citename{{Simpson} {\em et~al.\ }\relax,
  }2013]{2013MNRAS.429.2249S}
{Simpson}, F., {Heymans}, C., {Parkinson}, D., {Blake}, C., {Kilbinger}, M.,
  {Benjamin}, J., {Erben}, T., {Hildebrandt}, H., {Hoekstra}, H., {Kitching},
  T.~D., {Mellier}, Y., {Miller}, L., {Van Waerbeke}, L., {Coupon}, J., {Fu},
  L., {Harnois-D{\'e}raps}, J., {Hudson}, M.~J., {Kuijken}, K., {Rowe}, B.,
  {Schrabback}, T., {Semboloni}, E., {Vafaei}, S., \& {Velander}, M. 2013.
\newblock {CFHTLenS: testing the laws of gravity with tomographic weak lensing
  and redshift-space distortions}.
\newblock {\em \mnras}, {\bf 429}(Mar.), 2249--2263.

\bibitem[\protect\citename{{Spergel} {\em et~al.\ }\relax,
  }2013]{spergel2013afta}
{Spergel}, D., {Gehrels}, N., {Breckinridge}, J., {Donahue}, M., {Dressler},
  A., {Gaudi}, B.~S., {Greene}, T., {Guyon}, O., {Hirata}, C., {Kalirai}, J.,
  {Kasdin}, N.~J., {Moos}, W., {Perlmutter}, S., {Postman}, M., {Rauscher}, B.,
  {Rhodes}, J., {Wang}, Y., {Weinberg}, D., {Centrella}, J., {Traub}, W.,
  {Baltay}, C., {Colbert}, J., {Bennett}, D., {Kiessling}, A., {Macintosh}, B.,
  {Merten}, J., {Mortonson}, M., {Penny}, M., {Rozo}, E., {Savransky}, D.,
  {Stapelfeldt}, K., {Zu}, Y., {Baker}, C., {Cheng}, E., {Content}, D.,
  {Dooley}, J., {Foote}, M., {Goullioud}, R., {Grady}, K., {Jackson}, C.,
  {Kruk}, J., {Levine}, M., {Melton}, M., {Peddie}, C., {Ruffa}, J., \&
  {Shaklan}, S. 2013.
\newblock {Wide-Field InfraRed Survey Telescope-Astrophysics Focused Telescope
  Assets WFIRST-AFTA Final Report}.
\newblock {\em ArXiv e-prints}, May.

\bibitem[\protect\citename{{Spergel} {\em et~al.\ }\relax,
  }2015]{spergel2015wide}
{Spergel}, D., {Gehrels}, N., {Baltay}, C., {Bennett}, D., {Breckinridge}, J.,
  {Donahue}, M., {Dressler}, A., {Gaudi}, B.~S., {Greene}, T., {Guyon}, O.,
  {Hirata}, C., {Kalirai}, J., {Kasdin}, N.~J., {Macintosh}, B., {Moos}, W.,
  {Perlmutter}, S., {Postman}, M., {Rauscher}, B., {Rhodes}, J., {Wang}, Y.,
  {Weinberg}, D., {Benford}, D., {Hudson}, M., {Jeong}, W.-S., {Mellier}, Y.,
  {Traub}, W., {Yamada}, T., {Capak}, P., {Colbert}, J., {Masters}, D.,
  {Penny}, M., {Savransky}, D., {Stern}, D., {Zimmerman}, N., {Barry}, R.,
  {Bartusek}, L., {Carpenter}, K., {Cheng}, E., {Content}, D., {Dekens}, F.,
  {Demers}, R., {Grady}, K., {Jackson}, C., {Kuan}, G., {Kruk}, J., {Melton},
  M., {Nemati}, B., {Parvin}, B., {Poberezhskiy}, I., {Peddie}, C., {Ruffa},
  J., {Wallace}, J.~K., {Whipple}, A., {Wollack}, E., \& {Zhao}, F. 2015.
\newblock {Wide-Field InfrarRed Survey Telescope-Astrophysics Focused Telescope
  Assets WFIRST-AFTA 2015 Report}.
\newblock {\em ArXiv e-prints}, Mar.

\bibitem[\protect\citename{{Tinker} {\em et~al.\ }\relax,
  }2013]{2013ApJ...778...93T}
{Tinker}, J.~L., {Leauthaud}, A., {Bundy}, K., {George}, M.~R., {Behroozi}, P.,
  {Massey}, R., {Rhodes}, J., \& {Wechsler}, R.~H. 2013.
\newblock {Evolution of the Stellar-to-dark Matter Relation: Separating
  Star-forming and Passive Galaxies from z = 1 to 0}.
\newblock {\em \apj}, {\bf 778}(Dec.), 93.

\bibitem[\protect\citename{{Velander} {\em et~al.\ }\relax,
  }2014]{2014MNRAS.437.2111V}
{Velander}, M., {van Uitert}, E., {Hoekstra}, H., {Coupon}, J., {Erben}, T.,
  {Heymans}, C., {Hildebrandt}, H., {Kitching}, T.~D., {Mellier}, Y., {Miller},
  L., {Van Waerbeke}, L., {Bonnett}, C., {Fu}, L., {Giodini}, S., {Hudson},
  M.~J., {Kuijken}, K., {Rowe}, B., {Schrabback}, T., \& {Semboloni}, E. 2014.
\newblock {CFHTLenS: the relation between galaxy dark matter haloes and baryons
  from weak gravitational lensing}.
\newblock {\em \mnras}, {\bf 437}(Jan.), 2111--2136.

\bibitem[\protect\citename{{Wright} {\em et~al.\ }\relax, }2010]{WISE}
{Wright}, E.~L., {Eisenhardt}, P.~R.~M., {Mainzer}, A.~K., {Ressler}, M.~E.,
  {Cutri}, R.~M., {Jarrett}, T., {Kirkpatrick}, J.~D., {Padgett}, D.,
  {McMillan}, R.~S., {Skrutskie}, M., {Stanford}, S.~A., {Cohen}, M., {Walker},
  R.~G., {Mather}, J.~C., {Leisawitz}, D., {Gautier}, III, T.~N., {McLean}, I.,
  {Benford}, D., {Lonsdale}, C.~J., {Blain}, A., {Mendez}, B., {Irace}, W.~R.,
  {Duval}, V., {Liu}, F., {Royer}, D., {Heinrichsen}, I., {Howard}, J.,
  {Shannon}, M., {Kendall}, M., {Walsh}, A.~L., {Larsen}, M., {Cardon}, J.~G.,
  {Schick}, S., {Schwalm}, M., {Abid}, M., {Fabinsky}, B., {Naes}, L., \&
  {Tsai}, C.-W. 2010.
\newblock {The Wide-field Infrared Survey Explorer (WISE): Mission Description
  and Initial On-orbit Performance}.
\newblock {\em \aj}, {\bf 140}(Dec.), 1868--1881.

\bibitem[\protect\citename{{Zu} \& {Mandelbaum}, }2015]{2015MNRAS.454.1161Z}
{Zu}, Y., \& {Mandelbaum}, R. 2015.
\newblock {Mapping stellar content to dark matter haloes using galaxy
  clustering and galaxy-galaxy lensing in the SDSS DR7}.
\newblock {\em \mnras}, {\bf 454}(Dec.), 1161--1191.

\end{thebibliography}
\appendix
\section{Unweighted moments}
\label{app:unweighted_mom}
If the unweighted moments of the PSFs did not diverge, then it is possible to obtain an expression for the sizes of the PSFs with and without the effect of interpixel capacitance.
Eq.~\ref{eq:exact_unweighted_size} shows how the effective size of any profile increases as a function of the isotropic IPC parameters, which we now derive.
The crucial idea behind the derivation is the following: if we convolve two or more image profiles, their unweighted moments add, i.e.,
\begin{equation}
 {\bf M}_h = {\bf M}_f + {\bf M}_g
\end{equation}
where $f(\vec{x})$ and $g(\vec{x})$ are two image profiles, $h(\vec{x}) = (f\otimes g)(\vec{x})$ and ${\bf M}_f$ is the second moments of $f(\vec{x})$ and so on.

\emph{If} the object we are measuring is circularly symmetric and has a finite size $\sigma$ that can be defined from the unweighted moments, say like that of a Gaussian,
then it's moment matrix would be $\sigma^2 \mathbf{1}_2$, where $\mathbf{1}_2$ is the $2\times 2$ identity matrix.
The pixel response is given by a top-hat profile, whose second moments matrix given by $\frac{1}{12}\mathbf{1}_2$.
For the IPC kernel given in Eq.~\ref{eq:twopK}, the matrix of second moments is given by $(2\alpha+4\alpha') \mathbf{1}_2$.

Thus, the moment matrix of the image would be $(\sigma^2 + 1/12 + 2\alpha + 4\alpha')\mathbf{1}_2$ 
and the measured size of the PSF, as a function of $\alpha$ would be
\begin{equation}
 \sigma'(\alpha) = \sqrt{\sigma^2 + \frac{1}{12}+ 2\alpha+4\alpha'}
 \label{eq:forecast}
\end{equation}
If the moments of the object were given by a more generic $2\times 2$ symmetric matrix, say,
\begin{equation}
 M_G = \begin{pmatrix} \sigma^2_x & \epsilon \\ \epsilon & \sigma^2_y \end{pmatrix}
\end{equation}
then
\begin{equation}
 \sigma'(\alpha) = \sqrt{\sqrt{\sigma^2_x\sigma^2_y -\epsilon^2}+\frac{1}{12}+2\alpha+4\alpha'}
\end{equation}

Note that this equation does not depend on the profile of the PSF but only on the fact that the second moments do not diverge.
However, this expression is paricularly useful only for Gaussian light profiles since this expression also matches the adaptive size of such objects.
Moreover, the conditions that $\sigma \gg 1 \text{ pixel }$ and $\alpha, \alpha' \ll 1$ must be satisfied so that the non-Gaussianity introduced by the pixel response and the IPC kernel
is not significant.

\end{document}